\documentclass[12pt]{iopart}

\expandafter\let\csname equation*\endcsname=\relax
\expandafter\let\csname endequation*\endcsname=\relax
\usepackage{amsmath,amssymb,amsthm}

\usepackage{graphicx}
\usepackage{dcolumn}
\usepackage{bm}
\usepackage{subfigure}

\begin{document}


\title{Geometric phase and non-adiabatic resonance of the Rabi model}

\author{Sijiang Liu$^1$, Zhiguo L\"u$^{1,2,*}$ and Hang Zheng$^{1,2}$}

\address{$^1$ Key Laboratory of Artificial Structures and Quantum Control (Ministry of Education), School of Physics and Astronomy,\\ Shanghai Jiao Tong University, Shanghai 200240, China}
\address{$^2$ Collaborative Innovation Center of Advanced Microstructures, Nanjing University, Nanjing 210093, China}

\ead{zglv@sjtu.edu.cn}

\begin{indented}
\item[]Aug 2021
\end{indented}

\begin{abstract}
We investigate the effects of counterrotating terms on geometric phase and its relation to the resonance of the Rabi model. We apply the unitary transformation with a single parameter to the Rabi model and obtain the transformed Hamiltonian involving multiple harmonic terms. By combining the counter-rotating-hybridized rotating-wave method with time-dependent perturbation theory, we solve systematically time evolution operator and then obtain the geometric phase of the two-level system. Our results are beyond adiabatic approximation and rotating-wave approximation (RWA). Higher-order harmonic resonance happens when driving frequency is equal to higher-order subharmonic of the Rabi frequency. In comparison with numerically exact results, our calculated results are accurate over a wide range of parameters space, especially in higher-order harmonic resonance regimes. In these regimes we demonstrate geometric phases change dramatically while those of the RWA are smooth. The RWA is thoroughly invalid even if the driving strength is extremely weak. We find it is the higher-order harmonic terms that play an important role on the cyclic state and demonstrate the characters of geometric phase in higher-order harmonic resonance regime. We also present analytical formalism of the change rate of geometric phase and quasienergies, which agree well with numerically exact ones even in the strong driving case. The developed method can be applied to explore the dynamics of strongly driven qubits and physical properties of higher-order harmonic processes.
\end{abstract}

\section{\label{sec1}Introduction}



The geometric phase has been a subject of extensive study since its discovery by Berry \cite{berry1984,science,PRA55,PRA93}. Aharonov and Anandan generalized the concept of geometric phase from the periodic change of time-dependent quantum systems and removed adiabatic condition \cite{AA,AA1}. The Aharonov-Anandan (AA) geometric phase is further extended to non-unitary evolution. The geometric phase provides a unified description for a variety of effects in physics \cite{wang}. Via measurements of the geometric phase acquired by a two-level system, a route towards both large magnetic field range and high sensitivity has been introduced in Ref. \cite{shiyan2018} and the work shows that the geometric phase can be a versatile tool for quantum sensing applications. In open (hence non-Hermitian) systems, another extension of the geometric phase can be defined and leads to the concept of geometric dephasing. It has both dynamic and geometric origins and can either reduce or restore coherence even when no geometric phase is acquired \cite{shiyan2015}. Recently Abelian-geometric-phase-based nonadiabatic geometric one-qubit gates have been realized experimentally with a superconducting Xmon qubit, which is performed on two lowest levels of the Xmon qubit \cite{zhao2019experimental}.  In addition to fundamental interests, the geometric phase and its non-Abelian generalizations can form the basis of any quantum computation \cite{wang,PRA100}. Since such phases are immune to certain types of errors, in particular random fluctuations during time evolution, they are potential tools for robust quantum computation and quantum information processing. 

The AA phase is obtained by cyclic initial state $|\psi(0)\rangle$ defined as $| \psi(T)\rangle=e^{i\theta}| \psi(0)\rangle$ where $\theta$ is the total phase, $T$ a period of a time-dependent Hamiltonian $H(t)$ and $|\psi\rangle$ a cyclic state \cite{Moore}, which can be solved by evolution operator $ U(t)$. The eigenvectors of $U(T)$ are cyclic initial states so $U(T)$ directly determines cyclic state and geometric phase. The geometric phase is defined by subtracting the dynamical phase $\alpha$ from the total phase $\theta$
\begin{equation}
    \gamma=\theta-\alpha,
    \label{gamma}
\end{equation}
in which the dynamic phase $\alpha$ is defined as
\begin{equation}
    \alpha=-\int_{0}^{T}\langle\psi|H|\psi\rangle d\tau.
    \label{alpha}
\end{equation}
This decomposition illustrates the unique nature of the geometric phase \cite{AA}. In this paper geometric phase refers to AA phase.

The physics of driven quantum systems, as an attractive topic in quantum physics, has been widely studied for several decades \cite{qubit1,qubit2,qubit3,qubit4}.
At present, the importance of such systems is at the heart of quantum information processing. The prototype of driven quantum systems is the Rabi model (we set $\hbar=1$),
\begin{eqnarray}
    H(t) & = & \frac{\Delta }{2}{\sigma _z} + \frac{A}{2}\cos (\omega t){\sigma _x} \nonumber \\
    &=& \frac{\Delta }{2}{\sigma _z} + \frac{A}{4}( e^{-i\omega t}{\sigma _{+}}+e^ {i\omega t}{\sigma _{-}})
    + \frac{A}{4}(e^ {i\omega t}\sigma _{+}+e^ {-i\omega t}\sigma _{-}),
    \label{Rabi model}
\end{eqnarray}
where $\sigma_{x,y,z}$ is Pauli matrix and $\sigma_\pm=(\sigma_x\pm i\sigma_y)/2$. $\Delta$ is the transition frequency of the two-level system, $A$ and $\omega$  the amplitude and frequency of the linearly polarized driving field, respectively \cite{PhysRev.49.324}. If one drops the last terms, i.e., the counterrotating terms in Eq.(3), which is the well-known rotating-wave approximation (RWA), one could immediately solve exactly the dynamics of the Rabi-RWA model and find the famous Rabi oscillation \cite{book}. Recently, significant experiment development of strongly driven dynamics have been realized in the semiconductor devices and superconducting circuits, where the characteristic coupling is comparable with or larger than the transition frequencies. In some experiments, the dynamics of the qubit exhibits complex evolution when the driving strength approaches or exceeds the qubit transition frequency, which is a signature of the breakdown of the RWA. It is necessary to take into account the counterrotating interaction of driving field to demonstrate the dynamics of strongly driven systems.  Moreover, many interesting phenomena are attributed to the contribution of counterrotating terms, such as Bloch-Siegert shift, quantum Zeno effects, exotic dynamical phenomena \cite{dynamic2016,CHRW}. It is interesting and important to study how the counterrotating terms influence the dynamics and geometric phase in a wide range of parameter space, especially $\Delta > \omega$.

Over all of the approaches to driven physics, it is worth emphasizing the efficacy of the counter-rotating-hybridized rotating-wave (CHRW) method to analytically treat the driven dynamics and the counterrotating effects of the Rabi model \cite{CHRW}, which has been extensively utilized in many researches, such as dynamical evolution \cite{CHRW,dynamic2020}, Bloch-Siegert shifts \cite{BS2015,BS2016,BS2017}, resonance fluorescence \cite{2013} and so on. Recently, the CHRW method has been used to explore an optimal charging protocol by considering a time-dependent periodic classical drive, especially, a train of rectangular pulses, in quantum battery research \cite{dianchi2020}. Moreover, it also has been experimentally proved that counterrotating terms can produce relatively large and nonmonotonic Bloch-Siegert shifts in the cavity frequency as the system is driven through a quantum-to-classical transition \cite{BS2017}.  It is reasonable to infer that counterrotating terms certainly change the geometric phase of Rabi model. In this work, we will reveal that the AA phase of the Rabi model, which is calculated by the CHRW method taking counterrotating terms into consideration, is much different from that of the RWA.

The Rabi model whose detuning is larger than frequency is an important issue and presents interesting physical effects due to multiple harmonic processes, which mainly come from counterrotating terms. Since the RWA neglects counterrotating terms, it does not allow for the multiple harmonic processes. Therefore the effects of multiple harmonic processes are lack in the RWA case and we need the method beyond the RWA to consider the effects of multiple harmonic processes. By the combination of the CHRW method and perturbation theory, we clearly show the effects of multiple harmonic processes on the geometric phase, quasienergies and dynamics in much broad parameter regime. Moreover, we reveal analytically the stunning phenomenon of dramatic change of geometric phase, which happens in the resonance regime, due to higher-order harmonic processes. By comparison, the effects of higher-order harmonic processes on total phase are tiny.

In this work, we combine the CHRW method and perturbation theory to analytically calculate the geometric phase of the Rabi model. In the transformed Hamiltonian of the CHRW, a series of separate harmonic interaction terms are obtained\cite{CHRW}. The geometric phase of main resonance obtained by the CHRW Hamiltonian is accurate as well as the general tendency of off-resonance case in comparison with the numerically exact results. Its deviation from the RWA lucidly illustrates the effects of counterrotating terms on the geometric phase. On the other hand, in higher-order harmonic resonance regime, geometric phase changes dramatically while the results given by both the CHRW and RWA are quite smooth. This stunning phenomenon arises from higher-order harmonic interactions that have a significant role on cyclic initial state and geometric phase. Therefore, we apply perturbation theory based on the CHRW method to treat higher-order harmonic processes. Then we  obtain the accurate cyclic initial state, quasienergies and the gap between them in non-adiabatic resonance regime. It is feasible to analytically give the geometric phase of the 3rd harmonic resonance regime and those of higher-order harmonic resonance regime by the numerical calculation of perturbation theory. The developed method can be applied to study the effects of higher-order harmonic terms on the dynamics and geometric phase, especially in the higher-order harmonic resonance regime. The present research highlights the important effects of high-order harmonic terms on cyclic evolution and quasi-energies of non-adiabatic resonance. Moreover, dramatic change of geometric phase in higher-order harmonic resonance regimes widely exists in periodic quantum systems, such as the anisotropic Rabi model. The perturbative approach based on the CHRW method is still valid and can be extended to study those systems. Moreover, the present method can also be used to study the complicated dynamics of a strongly driven qubit.

The structure of the paper is organized as follows: In section \ref{sec2} we apply the CHRW method to show the effect of counterrotating terms on the geometric phase and we find in the main harmonic resonance regime the results of the CHRW are in good agreement with the numerically exact result but in the 3rd harmonic resonance regime the geometric phase of numerically exact method changes dramatically. Then in section \ref{sec3} we apply perturbation theory based on the CHRW method to take into account the effects of higher-order harmonic processes in order to calculate the geometric phase in the 3rd harmonic resonance regime. We demonstrate the effect of higher-order harmonic processes on geometric phase, quasienergy and dynamics. In section \ref{sec4} we apply perturbation theory to numerically calculate the geometric phase in higher-order harmonic resonance regime, for example, 5th harmonic resonance regime.  We demonstrate the effectiveness of perturbation theory based on the CHRW method in higher-order harmonic resonance regime. Finally, in section \ref{sec5} we give the conclusion of this paper.

\section{\label{sec2}Counter-rotating-hybridized rotating-wave method}
In order to analytically calculate geometric phase of the Rabi model, we perform the unitary transformation with a generator $S(t)=i\frac{A}{2\omega}\xi\sin(\omega t)\sigma_{x}$. Then, the transformed Hamiltonian \cite{CHRW}, $\tilde{H}(t)=e^{S(t)}H(t)e^{-S(t)}-ie^{S(t)}\frac{d}{dt}e^{-S(t)},$ can be readily written as
\begin{equation}
    \tilde{H}(t) = \frac{\Delta}{2}\left\{ \cos\left[\frac{A}{\omega}\xi\sin(\omega t)\right]\sigma_{z}+\sin\left[\frac{A}{\omega}\xi\sin(\omega t)\right]\sigma_{y}\right\}+\frac{A}{2}(1-\xi)\cos(\omega t)\sigma_{x}.
    \label{H_N}
\end{equation}
Using the identity
\begin{equation}
    \exp\left[i\frac{A}{\omega}\xi\sin(\omega t)\right]=\sum_{n=-\infty}^{\infty}J_{n}\left(\frac{A}{\omega}\xi\right)\exp(in\omega t),
\end{equation}
where $J_{n}(\cdot)$ represents the $n$th-order Bessel function
of the first kind, we divide the Hamiltonian into two parts $\tilde{H}(t)=H_{0}(t)+V(t)$,
\begin{eqnarray}\fl
    H_{0}(t)&=&\frac{\Delta}{2} J_{0}\left(\frac{A}{\omega}\xi\right)\sigma_{z}+\frac{A}{2}(1-\xi)\cos(\omega t)\sigma_{x}
    +\Delta J_{1}\left(\frac{A}{\omega}\xi\right)\sin(\omega t)\sigma_{y}\nonumber \\ \fl
    &=&\frac{\Delta}{2} J_{0}\left(\frac{A}{\omega}\xi\right)\sigma_{z}
    +\left[\frac{A}{4}(1-\xi)+\frac{\Delta J_{1}\left(\frac{A}{\omega}\xi\right)}{2}\right]( e^{-i\omega t}{\sigma _{+}}+e^ {i\omega t}{\sigma _{-}})\nonumber \\ \fl
    &&+\left[\frac{A}{4}(1-\xi)-\frac{\Delta J_{1}\left(\frac{A}{\omega}\xi\right)}{2}\right]( e^{i\omega t}{\sigma _{+}}+e^ {-i\omega t}{\sigma _{-}}),
\end{eqnarray}
\begin{eqnarray}\fl
    V(t) &= \sum_{n=1}^{\infty}H_{2n}(t)+H_{2n+1}(t)\nonumber\\\fl
    &= \Delta\sum_{n=1}^{\infty}J_{2n}\left(\frac{A}{\omega}\xi\right)\cos(2n\omega t)\sigma_{z}
    +\Delta \sum_{n=1}^{\infty}J_{2n+1}\left(\frac{A}{\omega}\xi\right)\sin\left[(2n+1)\omega t\right]\sigma_{y}.
    \label{VV}
\end{eqnarray}
%
%
$V(t)$ includes all higher-order harmonic terms ($n\geq 1$). To proceed, we determine $\xi$ by the following equation
\begin{equation}
    J_1\left(\frac{A}{\omega}\xi\right)\Delta=\frac{A}{2}(1-\xi)\equiv\frac{\tilde{A}}{4}.
    \label{eqxieq}
\end{equation}
From Eq. (\ref{eqxieq}) it is obvious to see that the parameter $\xi$ is dependent on both $\Delta/\omega$ and $A/\omega$ which are determined by physical quantities in some experiments. In our calculation, we solve self-consistently $\xi$ by Eq. (\ref{eqxieq}) with  a set of $\Delta/\omega$ and $A/\omega$.
Thus, $H_{0}(t)$ becomes
\begin{equation}
    H_{\rm{CHRW}}(t)=\frac{\tilde{\Delta}}{2}\sigma_{z}+\frac{\tilde{A}}{4}(e^{-i\omega t}\sigma_{+}+e^{i\omega t}\sigma_{-}),
\end{equation}
which holds the RWA-like interaction with the renormalized transition frequency  $\tilde{\Delta}=\Delta J_{0}\left(\frac{A}{\omega}\xi\right)$ and the renormalized driving strength $\tilde{A}$. Therefore, its evolution operator $U_{0}$ of the CHRW Hamiltonian can be analytically solved

\begin{equation}
    {U_0}(t) = \left( {\begin{array}{*{20}{c}}
    {{e^{ - i\frac{{\omega t}}{2}}}\left[ {\cos \left( {\frac{{\tilde \Omega t}}{2}} \right) - \frac{{i\tilde \delta }}{{\tilde \Omega }}\sin \left( {\frac{{\tilde \Omega t}}{2}} \right)} \right]}&{ - {e^{ - i\frac{{\omega t}}{2}}}\frac{{i\tilde A}}{{2\tilde \Omega }}\sin \left( {\frac{{\tilde \Omega t}}{2}} \right)}\\
    { - {e^{i\frac{{\omega t}}{2}}}\frac{{i\tilde A}}{{2\tilde \Omega }}\sin \left( {\frac{{\tilde \Omega t}}{2}} \right)}&{{e^{i\frac{{\omega t}}{2}}}\left[ {\cos \left( {\frac{{\tilde \Omega t}}{2}} \right) + \frac{{i\tilde \delta }}{{\tilde \Omega }}\sin \left( {\frac{{\tilde \Omega t}}{2}} \right)} \right]}
    \end{array}} \right),
    \label{U0}
\end{equation}
where $\tilde{\delta}=\tilde\Delta-\omega$ is the modified detuning, $\tilde\Omega=\sqrt{\tilde{\delta}^2+\frac{\tilde{A}^2}{4}}$ is the renormalized Rabi frequency.

The evolution operator $U(t)$ of original Hamiltonian Eq. (1) can be written as $U(t)=e^{-S(t)} \tilde{U}(t)$ by the evolution operator $\tilde{U}(t)$ of the transformed Hamiltonian $\tilde{H}(t)$. If neglecting the effects of $V(t)$ on $\tilde{U}(t)$, we straightforwardly obtain $U(t)\approx U_{\rm{CHRW}}(t)=e^{-S(t)}U_0(t)$. Since $S(T)=0$, the cyclic initial states of $H$ are the same as those of $\tilde{H}$. We get
\begin{equation}
    {U_0}(T) =  - \cos \left( {\frac{{\tilde \Omega T}}{2}} \right) + i\frac{{\tilde A}}{{2\tilde \Omega }}\sin \left( {\frac{{\tilde \Omega T}}{2}} \right){\sigma _x} + i\frac{{\tilde \delta }}{{\tilde \Omega }}\sin \left( {\frac{{\tilde \Omega T}}{2}} \right){\sigma _z},
    \label{U0T}
\end{equation}
and immediately obtain cyclic initial states
\begin{equation}
    |\pm\rangle=|\tilde{\pm}\rangle=\sqrt{\frac{2\tilde{\Omega}}{\tilde{\Omega}\mp\tilde{\delta}}}\left(\begin{array}{c}\frac{1}{2}\mp\frac{\tilde{\delta}}{2 \tilde{\Omega}} \\\mp\frac{\tilde{A}}{4\tilde{\Omega}}\end{array}\right) ,
    \label{chu}
\end{equation}
and corresponding eigenvalues $e^{-iq_\pm T}$ where $q_\pm$ are quasienergies
\begin{equation}
    q_\pm=\mp\frac{\tilde{\Omega}-\omega}{2}+n\omega.
    \label{q}
\end{equation}
The total phases are
\begin{equation}
    \theta_{\pm}=-q_\pm T=\pm\frac{ \tilde{\Omega}-\omega}{2}T.
    \label{total}
\end{equation}
Dynamic phases read
\begin{eqnarray}
    \alpha_{\pm} &=& -\int_{0}^{T}\langle\psi_\pm|H| \psi_\pm\rangle d\tau = -\int_0^T {\left\langle {\tilde  \pm } \right.\left| U_0^\dag{{e^S}H{e^{ - S}}}U_0 \right|\left. {\tilde  \pm } \right\rangle d\tau }\nonumber\\
    &=& \pm \frac{{\tilde A}}{{2\tilde \Omega }}\left( { \frac{{\tilde \Delta \tilde \delta }}{{\tilde A}} + \frac{A}{4} + \frac{{\tilde A}}{8}} \right)T,
    \label{dongli}
\end{eqnarray}
where $|\psi_\pm\rangle$ are cyclic states of $H$. Subtracting dynamical phase from overall phase by Eq. (\ref{gamma}),  we obtain geometric phase
\begin{eqnarray}\fl
    \gamma _ \pm  & = & \theta_\pm+\int_0^T \langle {\tilde  \pm } | U_0^\dag{{e^S}H{e^{ - S}}}U_0 | {\tilde  \pm } \rangle d\tau
    = \pm \left[ {\frac{{\tilde \Omega  - \omega }}{2} - \frac{{\tilde A}}{{2\tilde \Omega }}\left( {  \frac{{\tilde \Delta \tilde \delta }}{{\tilde A}} + \frac{A}{4} + \frac{{\tilde A}}{8}} \right)} \right]T.
\label{CHRW}
\end{eqnarray}
In contrast, geometric phases of the RWA are
\begin{equation}
    \gamma_{\pm}^{\rm{RWA}}=\pm\left( {1-\frac{{\delta}}{\Omega } } \right)\pi,
\label{RWA}
\end{equation}
where $\delta=\Delta-\omega$ and $\Omega=\sqrt{\delta^2+\frac{A^2}{4}}$. If setting $\xi=0$, one gets $\tilde{A}=2A$ and $\tilde\Delta=\Delta$.  By the replacements $\tilde{A}/2\to A$,  $\tilde\delta\to\delta$, and $\tilde\Omega\to\Omega$, Eq. (\ref{CHRW}) has the same mathematical form as Eq. (\ref{RWA}), which means the geometric phase of the CHRW method returns to that of the RWA by renormalization.  Besides, We also use Runge-Kutta techniques to obtain the numerically exact evolution of the Rabi model. The time step is restricted to some values in order to get accurate results of the evolution operator. Here the value of time evolution $U(t)$ is convergent when the time step $dt*\omega$ is $10^{-5}$. We verify the numerical errors defined by $|\rm det {(U^{\dag}(t)U(t))}-1|$ are much less than $10^{-12}$. Therefore, the comparison between the numerical results and our calculated results of the CHRW method illustrates the validity of the analytical formalism.

In the top panels of figures \ref{fig_A1} and \ref{fig_A2}, we show the geometric phases $\gamma_\pm$, which satisfy the complementary relation  $\gamma_{+} + \gamma_{-}=2\pi$, as a function of $\Delta/\omega$ including both main and 3rd harmonic resonance regimes for $A/\omega=1$ and $2$, respectively. We illuminate the main and 3rd harmonic resonance regime in the top panel of figure 1(b). The first intersection of the geometric phases, namely, $\gamma_{\pm}/\pi=1$, corresponds to the main resonance, while off-resonance indicates the regime away from the main resonance. Thus, the interval between two blue points which are corresponding to $\gamma=0.5 \pi$, are defined as the main resonance regime. Similarly, near $\Delta/\omega \sim 3$, the third harmonic resonance regime is defined as the interval of $\Delta/\omega$ between the black points which are corresponding to $\gamma=0.5 \pi$. For comparison, we also give the results of the RWA and numerically exact method. When $\Delta/\omega$ is relatively small, the geometric phases of the CHRW and numerically exact result tend to $0$ or $2\pi$ while the geometric phases of the RWA tend to nonzero. When $\Delta/\omega$ is much larger than 2, all geometric phases of the CHRW, RWA, and numerically exact methods tend to $0$ or $2\pi$, which is consistent with the adiabatic theorem, except for higher-order harmonic resonance regime. In the main harmonic resonance regime, the results of the CHRW are in good agreement with numerically exact results. The high-degree overlap between the two curves verifies the validity of the analytic result Eq. (\ref{CHRW}) in the main harmonic resonance regime. In contrast, the results of the RWA are quite different from the CHRW results, which becomes more explicit with increasing the driving strength $A/\omega$. Therefore counterrotating terms have a non-negligible effect on geometric phase in the main harmonic resonance regime.
An intersection between $\gamma_+$ and $\gamma_-$ happens in the main harmonic resonance regime. The value of $\Delta/\omega$ at this intersection is slightly less than 1 and decreases with increasing $A/\omega$. However, geometric phases of the RWA always precisely intersect at $\Delta/\omega=1$. The shift of the CHRW's intersection from the RWA's one indicates the influence of counterrotating terms on the main resonance. Moreover, the effect of counterrotating terms on geometric phase becomes manifest with increasing $A/\omega$.
In the vicinity of the 3rd harmonic resonance regime, the geometric phase changes dramatically and neither the CHRW nor RWA results are consistent with the numerically exact results. It is obvious to see that geometric phases in the 3rd harmonic resonance regime are approximately symmetric about the central point near $\Delta \sim 3 \omega$ which could be defined as the 3rd harmonic resonance point $\Delta_{\rm res}$. $\Delta_{\rm res}/\omega$ is uniquely determined by $A/\omega$. Actually, $\Delta_{\rm res}$ is close to $3\omega$, and its deviation from $3\omega$ increases with increasing $A/\omega$ as well as the width of the 3rd harmonic resonance regime. Thus a full understanding of 3rd harmonic resonance regime requires the incorporation of the higher-order harmonic terms in $V$ which is neglected in the CHRW method.

Next we show Rabi frequency $\tilde\Omega$ and quasienergies $q_\pm$ in the middle panels of figures \ref{fig_A1} and \ref{fig_A2}, respectively. Rabi frequency first drops and then rises through $2\omega$. When $\Delta/\omega > 1$, $\tilde\Omega/\omega$ is approximately proportional to $\Delta/\omega$, and the absolute value of the slope of $q_\pm/\omega$ is one half of that of $\tilde\Omega/\omega$. In the 3rd harmonic resonance regime, we find that $\tilde\Omega\approx2\omega$ and $q_\pm/\omega\approx\pm\frac{1}{2}$. In contrast, at the 3rd harmonic resonance point, Rabi frequency and quasienergies of the CHRW method can be exactly obtained $\tilde\Omega=2\omega$ and $q_\pm/\omega=\pm\frac{1}{2}$, respectively. However, the quasienergies of the numerically exact results have no intersection and there is a gap between quasienergies at the 3rd harmonic resonance point. Rabi frequency also shows a discontinuity at this point. It indicates that higher-order harmonic terms influence quasienergies and Rabi frequency in the 3rd harmonic resonance regime. Because of conservation of angular momentum, we give the non-adiabatic resonance condition of the Rabi model
\begin{equation}
    \tilde\Omega=2n\omega,
\end{equation}
which is approximately written as $\Delta\approx(2n+1)\omega$ for an extremely weak driving case.

In the bottom panels of figures \ref{fig_A1} and \ref{fig_A2} we show the squares of modules of the elements of the cyclic initial states as function of $\Delta/\omega$ for $A/\omega=1$ and $2$, respectively. We assume that one cyclic initial state
$\left|  +  \right\rangle  = \left( {\begin{array}{*{20}{c}}
{{c_1}}\\
{{c_2}}
\end{array}} \right)$. According to the orthogonality of the eigenvectors it is easy to get the other $\left|  -  \right\rangle  = \left( {\begin{array}{*{20}{c}}
{ - {c_2}}\\
{{c_1}}
\end{array}} \right)$ or $\left( {\begin{array}{*{20}{c}}
{{c_2}}\\
{ - {c_1}}
\end{array}} \right)$.
It is reasonable to show $|c_1|^2$ and $|c_2|^2$ of one cyclic initial state with $|c_1|^2+|c_2|^2=1$. We find that in the 3rd harmonic resonance regime, $|c_1|^2$ and $|c_2|^2$ also change rapidly and have an intersection. In fact, the intersection is slightly smaller than the 3rd harmonic resonance point. Even if geometric phase is related to cyclic initial state, the relation between them is complicated, which is shown in the \ref{appendix1}. In the 3rd harmonic resonance regime, the higher-order harmonic terms dominate the cyclic initial state, and determine the geometric phase.

\section{\label{sec3}Perturbation theory}
In the previous section, it is demonstrated that the CHRW method can give the accurate geometric phase of the main harmonic resonance regime due to proper inclusion of counterrotating terms. But it cannot predict the exact ones of the 3rd harmonic resonance regime because of the neglect of higher-order harmonic terms in $V$. Since the 3rd harmonic resonance regime is very narrow and elusive, it is hard to probe the dramatic change of geometric phase by an analytic method. In this section we combine the CHRW method with perturbation theory to reveal the subtle resonant picture by taking the 2nd and 3rd harmonic terms in $V$ as perturbation. We first calculate perturbation of evolution operator and then analytically obtain modified cyclic initial states and accurate geometric phases. At the same time, we explain the relation between geometric phase of 3rd harmonic resonance regime and the resonance condition by a transparent formalism. Moreover, we present the analytical results of change rate of geometric phase and gap between quasienergies at the 3rd harmonic resonance point.

\subsection{\label{sec3.1}Perturbation}
In the valid parameter regime of the CHRW method, the coefficients of higher-order harmonic terms $H_n$ ($n\geq2$) in Eq. (\ref{VV}) are much smaller than those of $H_{\rm CHRW}$. Thus, regarding the leading term $H_2$ in $V$ safely as perturbation, we apply perturbation theory to calculate $\tilde U(T)$. Now we divide the transformed Hamiltonian into two parts: unperturbed part $H_{\rm CHRW}$ and the perturbation $H^\prime$,
\begin{equation}
    \tilde{H} = {H_{\rm CHRW}} + \lambda H^\prime,
\end{equation}
and its corresponding time evolution operator
\begin{equation}
    \tilde{U} = {U_0} + \lambda U_1+ \cdots,
\end{equation}
where $i{\frac{d U_0}{dt}} = {H_{\rm CHRW}}{U_0}$, and a dimensionless parameter $\lambda$ is just used to keep track of the orders of perturbations. Since $\tilde{U}(t)$ satisfies the time-dependent Schr\"odinger equation $i\frac{d \tilde{U}}{dt} = \tilde{H}\tilde{U}$, we solve $U_1$. To first order
\begin{equation}
    i{{\frac{d U_{\rm{1}}}{dt} }} = {H_{\rm CHRW}}{U_{\rm{1}}} + H^\prime{U_0},
\end{equation}
then we analytically solve
\begin{equation}
    {U_{\rm{1}}(t)} =  - i{U_0}\int_0^t {U_0^{ - 1}H^\prime{U_0}d\tau}.
    \label{U1}
\end{equation}
In order to give the correction induced by the 2nd harmonic term
$H_2$, we require the integral part $\int_0^t{U_0^{ - 1}H_2{U_0}}d\tau$, namely,

\begin{eqnarray}
\fl
    \int_0^t {U_0^{ -1}H_2{U_0}d\tau} = - \Delta {J_{\rm{2}}}(Z)\frac{{\tilde A\tilde \delta }}{{2{{\tilde \Omega }^2}}}\left[\frac{{\tilde \Omega \sin (\tilde \Omega t)\cos (2\omega t) - 2\omega \cos (\tilde \Omega t)\sin (2\omega t)}}{{{{\tilde \Omega }^2} - 4{\omega ^2}}} \right.\nonumber \\\fl
    \left.- \frac{{\sin (2\omega t)}}{{2\omega }} \right]{\sigma _x}
    + \Delta {J_{\rm{2}}}(Z)\frac{{\tilde A}}{{2\tilde \Omega }}\frac{{\tilde \Omega  - \tilde \Omega \cos (\tilde \Omega t)\cos (2\omega t) - 2\omega \sin (\tilde \Omega t)\sin (2\omega t)}}{{{{\tilde \Omega }^2} - 4{\omega ^2}}}{\sigma _y}\nonumber \\\fl
    + \Delta {J_{\rm{2}}}(Z)\left[{\frac{{{{\tilde \delta }^2}}}{{{{\tilde \Omega }^2}}}\frac{{\sin (2\omega t)}}{{2\omega }} + \frac{{{{\tilde A}^2}}}{{4{{\tilde \Omega }^2}}}\frac{{\tilde \Omega \sin (\tilde \Omega t)\cos (2\omega t) - 2\omega \cos (\tilde \Omega t)\sin (2\omega t)}}{{{{\tilde \Omega }^2} - 4{\omega ^2}}}} \right]{\sigma _z},
    \label{H2yuan}
\end{eqnarray}
in which $Z=A\xi/\omega$.
Then letting the upper limit of the integral be $T$, we get
\begin{eqnarray}\fl
    \int_0^T {U_0^{ - 1}{H_2}{U_0}d\tau }\nonumber\\\fl
    = \frac{{\Delta {J_{\rm{2}}}(Z)\tilde A}}{{2({{\tilde \Omega }^2} - 4{\omega ^2})}}\left\{ { - \frac{{\tilde \delta }}{{\tilde \Omega }}\sin (\tilde \Omega T){\sigma _x}
    + \left[ {1 - \cos (\tilde \Omega T)} \right]{\sigma _y}
    + \frac{{\tilde A}}{{2\tilde \Omega }}\sin (\tilde \Omega T){\sigma _z}} \right\}.\nonumber\\
\end{eqnarray}
Finally, substituting Eq. (\ref{U0T}) into Eq. (\ref{U1}), we obtain the first order correction to $\tilde{U}$
\begin{eqnarray}\fl
    U_{1z}(T) &=& - i{U_0}(T)\int_0^T {U_0^{ - 1}{H_2}{U_0}d\tau }=\frac{{i\Delta {J_2}(Z)\tilde A\sin \left( {\frac{{\tilde \Omega T}}{2}} \right)}}{{{{\tilde \Omega }^2} - 4{\omega ^2}}}\left( { - \frac{{\tilde \delta }}{{\tilde \Omega }}{\sigma _x} + \frac{{\tilde A}}{{2\tilde \Omega }}{\sigma _z}} \right).
    \label{U1T2}
\end{eqnarray}
The whole time evolution operator $\tilde U(T) = {U_0}(T) + {U_{1z}}(T)$ yields
\begin{eqnarray} \fl
   \tilde U(T) &=& - \cos \left( {\frac{{\tilde \Omega T}}{2}} \right) + \frac{i}{{\tilde \Omega }}\sin \left( {\frac{{\tilde \Omega T}}{2}} \right)\left[ {\left( {\frac{{\tilde A}}{2} - k\tilde \delta } \right){\sigma _x} + \left( {\tilde \delta  + k\frac{{\tilde A}}{2}} \right){\sigma _z}} \right],
    \label{UTp}
\end{eqnarray}
where $k=\Delta J_2(Z)\tilde A/(\tilde\Omega^2-4\omega ^2)$.
We immediately obtain one cyclic initial state $\left|+\right\rangle$,
\begin{eqnarray}
    \left|  +  \right\rangle  = \frac{1}{L}\left( {\begin{array}{*{20}{l}}
    { - \tilde \delta  + \sqrt {1 + {k^2}} \tilde \Omega  - \frac{{k\tilde A}}{2}}\\
    { - \frac{{\tilde A}}{2} + k\tilde \delta }
    \end{array}} \right).
    \label{xinchutai}
\end{eqnarray}
where $L^{-1} =\left [\sqrt {2(1 + {k^2}){{\tilde \Omega }^2} - (2\tilde \delta  + k\tilde A)\sqrt {1 + {k^2}} \tilde \Omega }\right]^{-1}$ is the normalization factor. As the eigenvalues of $\tilde U(T)$ are very close to those of $U_0(T)$, total phases almost have no change. However, $U_{1z}(T)$ makes great contribution to $\tilde U(t)$. Since cyclic initial states are sensitive to evolution operator as well as $U_0(T)\approx-I$ (Identity Matrix), the eigenvectors of $\tilde U(T)$ are quite different from those of $U_0(T)$. Thus, both dynamic phase and geometric phase change sharply in the higher-order harmonic resonance regime. By Eqs. (\ref{gamma}) and (\ref{total}), we obtain the geometric phase corresponding to Eq. (\ref{xinchutai})
\begin{equation}
  \gamma_+  ={\theta _ + } + \int_0^T {\langle {\rm{ + }}|{{\tilde U}^\dag }(t){e^S}H{e^{ - S}}\tilde U(t)|{\rm{ + }}\rangle d\tau }.
\label{exact}
\end{equation}
In order to present the analytical character of geometric phase, here we use $U_0(t)$ instead of $\tilde U(t)$ in the integrand because the substitution of evolution operator leads to extremely small difference.
Besides, the other cyclic initial state $\left|-\right\rangle$ is orthogonal to $\left|+\right\rangle$.
Due to unitary property of $U$ and the orthogonality of the cyclic initial states, we get
\begin{equation}
    \left\langle  -  \right|{U^ \dag }HU\left|  -  \right\rangle=-\left\langle  +  \right|{U^ \dag }HU\left|  +  \right\rangle.
\end{equation}
Then $\alpha_-=-\alpha_+$. By $\theta_-=-\theta_+$, we get $\gamma_-=-\gamma_+$. Therefore, we obtain geometric phases
\begin{eqnarray}
    \gamma_\pm=\pm \left[ \frac{{\tilde \Omega  - \omega }}{2}T - \frac{1}{{\sqrt {1 + {k^2}} }}\left( {\frac{{\tilde \Delta \tilde \delta }}{{\tilde A}} + \frac{{\tilde A}}{8} + \frac{A}{4}} \right)\frac{{\tilde AT}}{{2\tilde \Omega }} \right.\nonumber\\
    \left.- \frac{k}{{\sqrt {1 + {k^2}} }}\left( {\frac{A}{4} - \frac{{\tilde A}}{8}} \right)\left( { - \tilde \delta  + 2\omega } \right)\frac{{\sin (\tilde \Omega T)}}{{{{\tilde \Omega }^2} - 4{\omega ^2}}} \right],
    \label{3HR}
\end{eqnarray}
which is derived in the \ref{appendix1}.

The 2nd harmonic term $H_2$ taken as perturbation contributes to the main correction of evolution operator. We have proved that our results are in good agreement with numerically exact results (shown in the next subsection). Nevertheless the results calculated by Eq. (\ref{UTp}) are not precise enough for a strong driving case. Consequently, we take the 3rd harmonic term into account, whose coefficient is an order of magnitude smaller than that of the 2nd harmonic term. Similarly, we obtain the first order perturbation induced by the 3rd harmonic term $H_{\rm{3}}=\Delta {J_{\rm{3}}}(Z)\sin (3\omega t){\sigma _y}$ under the condition of $\tilde\Omega\approx2\omega$
\begin{eqnarray}
    {U_{1y}}(T) = - \frac{{i\Delta {J_3}(Z)\left( {\tilde \Omega  + \tilde \delta } \right)\sin \left( {\frac{{\tilde \Omega T}}{2}} \right)}}{{{{\tilde \Omega }^2} - 4{\omega ^2}}}\left( { - \frac{{\tilde \delta }}{{\tilde \Omega }}{\sigma _x} + \frac{{\tilde A}}{{2\tilde \Omega }}{\sigma _z}} \right),
    \label{U1T3}
\end{eqnarray}
which is derived in the \ref{appendix1}. It is noticed that the first order perturbation induced by $H_3$ have a similar mathematical structure as that of $H_2$ and the only difference between them is the distinct coefficients of  ${U_{1z}}(T)$ and ${U_{1y}}(T)$. Combining ${U_{1z}}(T)$ in Eq.(\ref{U1T2}) and ${U_{1y}}(T)$ in Eq.(\ref{U1T3}) together, we get the total first order correction to $\tilde{U}$
\begin{eqnarray}
    {U_1}(T) &= & {U_{1z}}(T) + {U_{1y}}(T)=ik\sin \left( {\frac{{\tilde \Omega T}}{2}} \right)\left( { - \frac{{\tilde \delta }}{{\tilde \Omega }}{\sigma _x} + \frac{{\tilde A}}{{2\tilde \Omega }}{\sigma _z}} \right),
    \label{total U1}
\end{eqnarray}
where $k$ is rewritten as
\begin{equation}
    k = \frac{{\Delta \left[ {{J_2}\left( Z \right)\tilde A - {J_3}\left( Z \right)\left( {\tilde \Omega  + \tilde \delta } \right)} \right]}}{{{{\tilde \Omega }^2} - 4{\omega ^2}}}.
    \label{k}
\end{equation}
Only replacing $k$ in Eqs. (\ref{xinchutai}) and (\ref{3HR}) with Eq. (\ref{k}), we obtain the modified cyclic initial states and corresponding geometric phases. Since ${U_{1y}}(T)$ is an order of magnitude smaller than ${U_{1z}}(T)$, ${U_{1y}}(T)$ makes a little contribution to the time evolution and does not change the main trend of geometric phase. After taking the perturbation induced by the next leading term $H_3$ into account additionally, we prove that the perturbative results are in perfect agreement with numerically exact results. If $k=0$, Eqs. (\ref{UTp}), (\ref{xinchutai}) and (\ref{3HR}) have the same forms as Eqs. (\ref{U0T}), (\ref{chu}) and (\ref{CHRW}), respectively. It indicates that the results of the CHRW method is the limit case of those of the perturbation theory based on the CHRW method.

\subsection{\label{sec3.2}Results and discussion}
In figure \ref{fig_PT}, we show the geometric phases of perturbation theory based on the CHRW method, which is calculated by Eq. (\ref{3HR}).  The CHRW and numerically exact results are also depicted for comparison.
By the resonance condition $\tilde\Omega=2\omega$, 3rd harmonic resonance point $\Delta_{\rm res}$ can be expanded up to second order in $A/\omega$
\begin{equation}
    \Delta_{\rm res}\approx\left( {3 - \frac{{3{A^2}}}{{32{\omega ^2}}}} \right)\omega,
    \label{point}
\end{equation}
which is given in the \ref{appendix2}. This shift $\frac{3A^2}{32\omega }$ in Eq. (\ref{point}) is consistent with Bloch-Siegert shift of 3rd harmonic resonance in Ref. \cite{Shirley}.

The perturbation theory based on the CHRW method gives the accurate geometric phase in much a broader parameter regime. In the top panels of figures \ref{fig_A1PT} and \ref{fig_A2PT} we show geometric phases given by Eq. (\ref{3HR}), those of Eq. (\ref{CHRW}) and numerically exact results in both main and 3rd harmonic resonance regimes for $A/\omega=1$ and $2$, respectively. In the main harmonic resonance regime the perturbative results given by Eq. (\ref{3HR}) returns to those of Eq. (\ref{CHRW}). It is obvious to see that Eq. (\ref{3HR}) is valid in both main and 3rd harmonic resonance regimes. In the top panels of figures \ref{fig_D3A1} and \ref{fig_D3A2}, we zoom in on the geometric phases in the 3rd harmonic resonance regime for $A/\omega=1$ and $2$, respectively. The high-degree overlap between Eq. (\ref{3HR}) and numerically exact result verifies the validity of Eq. (\ref{3HR}). In contrast, geometric phases of the CHRW method are smooth and the difference from the other results indicates that it is the effect of higher-order harmonic terms give rise to the sharp change of geometric phase. Geometric phases $\gamma_+$ and $\gamma_-$ intersect at $\pi$ for three times, including the 3rd harmonic resonance point. The absolute value of $k$ represents the intensity of effects of higher-order harmonic terms, which is large in the 3rd harmonic resonance regime. In contrast, owing to an extremely small value of $k$ in the main harmonic resonance regime we confirm that higher-order harmonic terms have negligible effects on geometric phase for $\Delta<2\omega$.

A dramatic change of geometric phase happens in the 3rd harmonic resonance regime and it can be described by our analytical results. The second term in Eq. (\ref{3HR}) dominates the change of geometric phase in the 3rd harmonic resonance regime. Near the 3rd harmonic resonance point, geometric phase satisfies approximately a linear relation with $\Delta/\omega$. Moreover,
the dimensionless change rate of geometric phase $\left|\frac{d\gamma}{d(\Delta/\omega)}\right|$ can be expanded to third order in $\omega/A$
\begin{equation}
    \left| {\frac{{d\gamma }}{{d(\Delta /\omega )}}} \right|= 3\pi \left[ {128{{\left( {\frac{\omega }{A}} \right)}^3} - \frac{{63}}{8} {\frac{\omega }{A}}} \right],
    \label{xielv}
\end{equation}
which is derived in the  \ref{appendix2.1}. In figure \ref{fig_xielv}, we show the scaled ratio $\frac{1}{100 ~\omega/A}\left|\frac{d\gamma}{d(\Delta/\omega)}\right|$ as a function of $(\omega/A)^2$. For comparison, we also give the numerically exact result. Obviously, the result calculated by Eq. (\ref{xielv}) is in good agreement with numerically exact one and both of them are linearly dependent of $(\omega/A)^2$. At the same time, the great rate of the change of geometric phase illustrates the sensitivity to the energy splitting $\Delta$. Therefore, the analytic result Eq. (\ref{xielv}) demonstrates that there happens a dramatic change of geometric phase as a function of driving parameter in the 3rd harmonic resonance regime.

From Eq. (\ref{UTp}), quasienergies $q_\pm$ are obtained under the condition of $\tilde\Omega\approx2\omega$
\begin{eqnarray}
    {q_ \pm } & = & \mp \frac{{\arctan \left[ {\sqrt {{k^2} + 1} \tan \left( {\frac{{\tilde \Omega T}}{2}} \right)} \right]}}{T} + \frac{{\left( {2n + 1} \right)\omega }}{2} \nonumber\\
    & \approx & \mp \sqrt {{k^2} + 1} \frac{{\tilde \Omega  - 2\omega }}{2} + \frac{{\left( {2n + 1} \right)\omega }}{2},
    \label{quasienergy}
\end{eqnarray}
and the modified Rabi frequency given by  perturbation theory is written as
\begin{equation}
    {{\tilde \Omega }_p} = \sqrt {{k^2} + 1} (\tilde \Omega  - 2\omega ) + 2\omega.
    \label{O_p}
\end{equation}
In the middle panels of figure \ref{fig_PT} we show quasienergies and Rabi frequency given by Eqs. (\ref{quasienergy}) and (\ref{O_p}), those of Eq. (\ref{q}) and numerically exact results in both main and 3rd harmonic resonance regimes. In the main harmonic resonance regime quasienergies given by Eq. (\ref{quasienergy}) returns to those given by Eq. (\ref{q}). It is obvious to see that Eq. (\ref{quasienergy}) is valid in both main and 3rd harmonic resonance regimes.
In the middle panels of figures \ref{fig_D3A1} and \ref{fig_D3A2}, we zoom in on the quasienergies $q_\pm$ in the 3rd harmonic resonance regime for $A/\omega=1$ and $2$, respectively.
The perfect overlap between the results of perturbation theory and numerically exact results verifies the validity of Eq. (\ref{quasienergy}). It is clearly to see that there exists a gap for quasienergies of both perturbation theory and numerically exact results at the 3rd harmonic resonance point. In contrast, quasienergies in Eq. (\ref{q}) intersect at 3rd harmonic resonance regime and this subtle difference from the other results indicates the significant effect of higher-order harmonic terms gives rise to the evident gap between quasienergies.

An avoided crossing of quasienergies happens at the 3rd harmonic resonance point and there is a gap $\Xi$ between quasienergies. The dimensionless gap $\Xi/\omega$ can be expanded up to fifth order in $A/\omega$
\begin{equation}
    \frac{\Xi}{\omega}= \frac{{{A^3}}}{{128{\omega ^3}}}\left( {1 + \frac{{17{A^2}}}{{2048{\omega ^2}}}} \right),
    \label{gap}
\end{equation}
which is derived in the  \ref{appendix2.2}. In figure \ref{fig_gap}, we show the dimensionless gap $\Xi/\omega$ as a function of $A/\omega$. It is obvious to see the good agreement of Eq. (\ref{gap}) with numerically exact result. For comparison, we also give the result of $(A/\omega)^3/128$, whose curve overlaps numerically exact result. Therefore, we obtain a simple leading order relationship $\Xi/\omega\propto(A/\omega)^3$. Near the 3rd harmonic resonance point, from detailed analysis in the \ref{appendix2.2}, we prove that quasienergies satisfy
\begin{equation}
    q_\pm\pm\left[\frac{\Xi}{2}+\frac{(2n+1)\omega}{2}\right]\propto (\Delta-\Delta_{\rm res})^2.
    \label{q zhengbi}
\end{equation}

In the bottom panels of figure \ref{fig_PT} we show the squares of modules of the elements of the cyclic initial states given by Eq. (\ref{xinchutai}), those of Eq. (\ref{chu}) and numerically exact results in both main and 3rd harmonic resonance regimes. In the main harmonic resonance regime cyclic initial states of perturbation theory return to cyclic initial states of the CHRW method. It is obvious to see that Eq. (\ref{xinchutai}) is valid in both main and 3rd harmonic resonance regimes. In the bottom panels of figures \ref{fig_D3A1} and \ref{fig_D3A2}, we zoom in on the squares of modules of the elements of the cyclic initial states in the 3rd harmonic resonance regime for $A/\omega=1$ and $2$, respectively. The results of perturbation theory overlap numerically exact results, which verifies the validity of Eq. (\ref{xinchutai}). In contrast, the results of the CHRW method are smooth and flat. The difference from the other results illustrate it is the effect of higher-order harmonic terms that leads to the sharp change of cyclic initial states. The squares of modules of the elements of the cyclic initial states intersect for one time. The value of $\Delta/\omega$ corresponding to the intersection is slightly smaller than the 3rd harmonic resonance point $\Delta_{\rm res}$.

All results of perturbation theory are valid in both main and 3rd harmonic resonance regimes. If $k=0$, they reduce to the results of the CHRW method. In perturbation theory a dimensionless quantity $k$ is a signature, which is the coefficient of combined perturbation in Eq. (\ref{total U1}). The absolute value of $k$ demonstrates the intensity of effects of higher-order harmonic terms. There is a positive correlation between effects of higher-order harmonic terms and the absolute value of $k$. In the main harmonic resonance regime, due to $|k|\ll1$  effects of higher-order harmonic terms is neglectable.  Consequently, the CHRW method is valid. While in the 3rd harmonic resonance regime $k$ becomes larger so that effects of higher-order harmonic terms become important. Thus, the perturbation calculation based on the CHRW method is necessary. Especially when the absolute value of $k$ is infinity at the 3rd harmonic resonance point, the effects of higher-order harmonic terms are strongest.

Higher-order resonance satisfies the commonly used adiabatic approximation criterion
\begin{equation}
    \left|\frac{\langle\phi_{1}|\dot H|\phi_{2}\rangle}{(E_2-E_1)^{2}}\right| \ll 1,
\end{equation}
where $\phi_{i}(i=1,2)$ are one instantaneous eigenstate of $H$ and $E_{1,2}$ are their corresponding eigenvalues.
However, instantaneous eigenstates are totally different from cyclic states because of effects of higher-order harmonic terms. Berry phase is geometric phase for adiabatic evolution while AA phase is geometric phase for cyclic evolution. In the 3rd harmonic resonance regime the Berry phase of Rabi model are completely different from AA phase. It indicates that instantaneous eigenstate is different from cyclic state in this regime.
This demonstrates not all physical properties in the adiabatic limit should be obtainable from the instantaneous eigenstate. Therefore, we have to consider the effect of higher-order harmonic terms on cyclic states.

Higher-order harmonic processes lead to higher-order harmonic resonance which have important effects on cyclic state and its evolution. The cyclic state is different from the instantaneous eigenstate of the Rabi model in the higher-order harmonic resonance regime. In figure \ref{fig4}, we show the time evolution of the two-level system which is initially prepared as instantaneous eigenstate or cyclic state in the 3rd harmonic resonance regime, on the Bloch sphere for $\Delta/\omega=2.9$ and $A/\omega=1$. It is obvious to see in figure \ref{fig4}(a), after one period evolution of the case initially prepared as instantaneous eigenstate, the final state almost returns to the initial one. In contrast, after a long-time evolution shown in figure \ref{fig4}(b), the state of the two-level system gradually deviates from the trajectory of figure \ref{fig4}(a) and the final state at $t/T=10$ is much different from the initial state. With increasing evolution time, the difference between the final and initial states becomes larger and the periodicity is broken. It turns out that the adiabatic condition is invalid for the case initially prepared as instantaneous eigenstate in the resonance regime. For the case away from the 3rd harmonic resonance (not shown here), the close trajectory on the Bloch sphere happens. It indicates the adiabatic condition is valid away from the higher-order harmonic resonance. By comparison, it is obvious to see that the results of cyclic state in figures \ref{fig4}(c) and \ref{fig4}(d) show perfect cyclic evolution which is totally different from those of instantaneous eigenstate. Whether evolution time is short or long, the final state at $t=n T (n=1,2,3...)$ is exactly the same as the initial state with a total phase in the cases of both near 3rd harmonic resonance and away from 3rd harmonic resonance. These results intuitively verify that the evolution of the cyclic state is totally different from that of the instantaneous state in the higher-order harmonic resonance regime.

\section{\label{sec4}General perturbation in higher-order harmonic resonance regime}
In this section we apply perturbation theory based on the CHRW method to give the geometric phases in the higher-order harmonic resonance regime, for example, 5th harmonic resonance regime.
For general perturbations of higher-order harmonic terms, we replace $H^\prime$ with $V$ in perturbation theory. Setting $\tilde U=U_0W$, we use Schr\"odinger equation $i{\frac{d U_0}{dt}} = {H_{\rm CHRW}}{U_0}$ and $i\frac{d}{dt}\tilde U = \tilde H\tilde U = \left( {{H_{\rm CHRW}} + V} \right)\tilde U$ to obtain
\begin{equation}
    i \frac{d W}{dt} =  {H_V}W,
\end{equation}
where $H_V=U_0^{-1}VU_0$. From above equation $W$ can be expressed in the series form
\begin{eqnarray}
    W & = & I + ( - i)\int_0^t {d{\tau _1}{H_V}} + {( - i)^2}\int_0^t {d{\tau _1}\int_0^{{\tau _1}} {d{\tau _2}{H_V}({\tau _1}){H_V}({\tau _2})} }  +  \cdots .
\end{eqnarray}

First and second order corrections of second to fifth harmonic terms are considered
\begin{equation}
    {W_n} =  - i\int_0^t {d{\tau _1}U_0^{ - 1}{H_n}{U_0}} ,
\end{equation}
\begin{equation}
    {W_{nm}}={( - i)^2}\int_0^t {d{\tau _1}U_0^{ - 1}{H_n}{U_0}\int_0^{{\tau _1}} {d{\tau _2}U_0^{ - 1}{H_m}{U_0}}} ,
\end{equation}
where $n,m\geq2$. Since $\tilde\Omega\approx4\omega$, $W_2(T)$ is not the leading term in $W(T)$ though the coefficient of the term $H_2$ is the largest in $V$. Thus, it is necessary to consider the first order and second order corrections of other terms in $V$ near the 5th harmonic resonance regime. $W(T)$ is approximately obtained
\begin{equation}
    W(T)\approx I+W_{22}(T)+W_{3}(T)+W_{4}(T)+W_{23}(T)+W_{32}(T).
    \label{W}
\end{equation}
$W_{22}(T)$ and $W_3(T)$ make main contribution to $W(T)$ and they dominate the trend of geometric phases in the 5th harmonic resonance regime. Furthermore, $W_4(T)$, $W_{23}(T)$ and $W_{32}(T)$ modify the geometric phase slightly. The main term  $W_{22}(T)$ comes from the interaction of 2nd harmonic term and itself. In the 5th harmonic resonance regime, we should not only consider the effects of single harmonic term, but also the effects of cross interactions of harmonic terms.

We use numerical integration to calculate Eq. (\ref{W}). In the top panel of figure \ref{fig_D5} we show the geometric phases obtained by Eq. (\ref{W}) and numerically exact results. The high-degree overlap between two curves verifies the validity of perturbation theory based on the CHRW method in the 5th harmonic resonance regime. Geometric phases also change dramatically and have five intersections, including the 5th harmonic resonance point. In the middle panel of figure \ref{fig_D5} we show quasienergies obtained by Eq. (\ref{W}) and numerically exact results. It proves that the terms taken into account in Eq. (\ref{W}) could give the accurate results of quasienergies. Similarly, there is also a gap between quasienergies at the 5th harmonic resonance point. In the bottom panel of figure \ref{fig_D5} we show the squares of modules of the elements of cyclic initial state obtained by Eq. (\ref{W}) and numerically exact results. Cyclic initial states in the 5th harmonic resonance regime are similar to those in the 3rd harmonic resonance regime. In comparison with the 3rd harmonic resonance treatment, it is necessary to take into account the interactions of higher-order harmonic terms to illustrate the features of higher-order harmonic resonance. Perturbation theory based on the CHRW method is valid in higher-order harmonic resonance regime. The width of higher-order harmonic resonance regime becomes much narrower and the interactions between higher-order harmonic terms become more important and complex.

\section{\label{sec5}Conclusion}
In this work, we combine the CHRW method and perturbation theory to systematically investigate the geometric phase of the Rabi model beyond the RWA. In the main harmonic resonance regime geometric phase obtained by the CHRW method is consistent with the numerically exact result. In higher-order harmonic resonance regime cyclic initial state and geometric phase changes dramatically with the increase of driving parameters and there exists a gap between quasienergies at any higher-order harmonic resonance point. In higher-order harmonic resonance regime the results obtained by perturbation theory based on the CHRW method are in perfect agreement with the numerically exact results. For example, we precisely calculate geometric phase in the 3rd harmonic resonance regime and clearly present some interesting and important characters: (i) since cyclic initial states are sensitive to $U(T)$ and $U(T)\approx-I$, the perturbations of higher-order harmonic terms have a huge influence on cyclic initial states; (ii) the dramatic change of cyclic initial states leads to the sharp change of geometric phases; (iii) there is a gap between quasienergies at the 3rd harmonic resonance point. Therefore, perturbation theory based on the CHRW method does give accurate time evolution operator to calculate accurate geometric phase and further analyse dynamical features in the Rabi model.
As the higher-order harmonic processes lead to non-adiabatic resonance, they have crucial influence on the geometric phase and quasienergies in the higher-order harmonic resonance regime. We predict the similar effects of higher-order harmonic processes happen in other periodic systems.

Since the nonadiabatic resonance happens under the condition of $\tilde\Omega=2n\omega$, there exists the sharp change of geometric phase near $\Delta=(2n+1)\omega$. The geometric phase can be measured by the population and detected in Ref. \cite{science}, and the detuning is realized in the Nitrogen-Vacancy spin experiment \cite{wood}. Since the higher-order harmonic resonance regime is very narrow and elusive, it is important in experiments to match the nonadiabatic resonance condition corresponding to the dramatic change of geometric phase and it requires the highly precise and stable driving field.
Theoretical findings of this work pave the way for experimentally probing the properties of periodically driven system with geometric phases. The geometric phases in the 3rd harmonic resonance regime might be measured through a quantum simulator, such as Floquet Raman system \cite{PhysRevLett.121.210501}. The dramatic change of geometric phase in higher-order harmonic resonance regime can be used to measure high-accuracy magnetic field. On the other hand, the magnetic field high accuracy can be measured using maser technology in Ref. \cite{jiang2021floquet} and geometric phase in the 3rd harmonic resonance regime may be detected.

The applications of the perturbative approach based on the CHRW Hamiltonian to a variety of realistic problems are desirable. In this paper, our results indicate the cyclic evolution is essentially different from adiabatic evolution and the RWA is totally invalid in the resonance regime, even if driving strength is much less than driving frequency. Besides, the developed method can be applied to explore complicated dynamics of a strongly driven qubit. We predict dramatic change of geometric phase and higher-order harmonic resonance widely exist in the periodic quantum systems. The combination of CHRW method and perturbation theory is a general method to explore the effects of higher-order harmonic processes, especially in higher-order harmonic resonance cases.

\ack
The work is supported by National Natural Science Foundation of China (Grants No. 11774226, No. 11874260, and No. 61927822).

Data availability statement

All data that support the findings of this study are included within the article (and any
supplementary files)

\appendix

\section{The perturbation theory\label{appendix1}}

\subsection{Perturbation induced by the 3rd harmonic term\label{appendix1.1}}
We define
\begin{equation}
    U_0=\left( {\begin{array}{*{20}{c}}
    {{U_{11}}}&{{U_{12}}}\\
    {{U_{21}}}&{{U_{22}}}
    \end{array}} \right).
    \label{1212}
\end{equation}
Then, we first calculate  $U_0^{ - 1}{H_3}{U_0}$
\begin{eqnarray}
\fl
    U_0^{ - 1}{H_3}{U_0} & = & \Delta J_3(Z) \sin \left( {3\omega t} \right){U_0}^{ - 1}{\sigma _y}{U_0}\nonumber\\\fl
    & = & i\Delta J_3(Z) \sin \left( {3\omega t} \right)\left( {\begin{array}{*{20}{c}}
    { - {U_{11}}{U_{12}} - {U_{21}}{U_{22}}}&{ - {U_{12}}^2 - {U_{22}}^2}\\
    {{U_{11}}^2 + {U_{21}}^2}&{{U_{11}}{U_{12}} + {U_{21}}{U_{22}}}
    \end{array}} \right).
\end{eqnarray}
By Eq. (\ref{U0}) the elements in the matrix above can be written as
\begin{eqnarray}\fl
    -({U_{11}}{U_{12}} + {U_{21}}{U_{22}}) & = &  \frac{{i\tilde A}}{{2\tilde \Omega }}
    \left\{\cos (\omega t)\sin (\tilde \Omega t)-\frac{{\tilde \delta }}{{\tilde \Omega }}\sin (\omega t)\left[ {1 - \cos (\tilde \Omega t)} \right]\right\},
\end{eqnarray}
\begin{eqnarray}\fl
    {U_{11}}^2 + {U_{21}}^2 =\left[ {\cos (\omega t) - i\frac{{{{\tilde \delta }^2}}}{{{{\tilde \Omega }^2}}}\sin (\omega t)} \right]\cos (\tilde \Omega t) - i\left( {1 - \frac{{{{\tilde \delta }^2}}}{{{{\tilde \Omega }^2}}}} \right)\sin (\omega t) - i\frac{{\tilde \delta }}{{\tilde \Omega }}{e^{ - i\omega t}}\sin (\tilde \Omega t).\nonumber\\
\end{eqnarray}
Then letting the upper limit of the integral  $\int_0^t U_0^{ - 1}{H_3}{U_0}$ be $T$ we get integrals of the elements.
When $\tilde\Omega\approx2\omega$ in the 3rd harmonic resonance regime, the terms whose frequencies are close to zero contribute to the main part of integrals. Thus, we obtain
\begin{eqnarray}
    \fl\int_0^T {\sin \left( {3\omega \tau } \right)\left( { - {U_{11}}{U_{12}} - {U_{21}}{U_{22}}} \right)d\tau }  \nonumber\\
    \fl\approx \frac{{i\tilde A}}{{4\tilde \Omega }}\int_0^T {\left[ {\sin (2\omega\tau )\sin (\tilde \Omega\tau ) + \frac{{\tilde \delta }}{{\tilde \Omega }}\cos (2\omega\tau )\cos (\tilde \Omega\tau )} \right]d\tau }
    \approx\frac{{i\tilde A(\tilde \Omega  + \tilde \delta )}}{{4{{\tilde \Omega }^2}(\tilde \Omega  - 2\omega )}}\sin \left(\frac{{\tilde \Omega T}}{2}\right),
\end{eqnarray}
\begin{eqnarray}
    \int_0^T {\sin \left( {3\omega \tau } \right)\left( {U_{11}^2 + U_{21}^2} \right)d\tau } \nonumber\\
    \approx \int_0^T \left[  - \frac{i}{2}\frac{{{{\tilde \delta }^2}}}{{{{\tilde \Omega }^2}}}\cos (2\omega \tau )\cos (\tilde \Omega \tau )
    - \frac{1}{2}\frac{{i\tilde \delta }}{{\tilde \Omega }}\sin (2\omega \tau )\sin (\tilde \Omega \tau ) \right]d\tau \nonumber\\
    \approx - \frac{{i\tilde \delta (\tilde \Omega  + \tilde \delta )}}{{2{{\tilde \Omega }^2}(\tilde \Omega  - 2\omega )}}\sin \left(\frac{{\tilde \Omega T}}{2}\right).\nonumber\\
\end{eqnarray}
Finally, by the approximation $U_0(T)\approx-I$ we get Eq. (\ref{U1T3})
\begin{eqnarray}
    {U_{1y}}(T) &=& - i{U_0}(T)\int_0^T {U_0^{ - 1}{H_3}{U_0}d\tau}  \approx i\int_0^T {U_0^{ - 1}{H_3}{U_0}d\tau}\nonumber\\
    &=& - \frac{{i\Delta {J_3}(Z)\left( {\tilde \Omega  + \tilde \delta } \right)\sin\left(\frac{\tilde \Omega T}{2}\right)  }}{{\tilde \Omega \left( {{{\tilde \Omega }^2} - 4{\omega ^2}} \right)}}\left( {-\tilde \delta {\sigma _x} + \frac{{\tilde A}}{2}{\sigma _z}} \right).\nonumber\\
\end{eqnarray}

\subsection{Geometric phase in the 3rd harmonic resonance regime\label{appendix1.2}}
We define one cyclic state as $\tilde\psi_+$ which satisfies $i\frac{d\tilde\psi_+}{dt} = \tilde H\tilde\psi_+$ and $\tilde\psi_+(0) = \left|  +  \right\rangle=\left( \begin{array}{l}{c_1}\\{c_2}\end{array} \right)$ in Eq. (\ref{xinchutai}).
Since ${e^S}H{e^{ - S}} \approx \frac{{\tilde \Delta }}{2}{\sigma _z} + \frac{{\tilde A}}{4}{\rm{sin}}(\omega t){\sigma _y} + \frac{A}{2}\cos (\omega t){\sigma _x}$, the dynamic phase $\alpha_+$ can be written as
\begin{eqnarray}
    \alpha_+&=&\int_0^T {\langle {{\tilde \psi }_ + }|{e^S}H{e^{ - S}}|{{\tilde \psi }_ + }\rangle d\tau }  = \frac{{\tilde \Delta }}{2}\int_0^T {\langle {{\tilde \psi }_ + }|{\sigma _z}|{{\tilde \psi }_ + }\rangle d\tau }   \nonumber\\
    && + \frac{{\tilde A}}{4}\int_0^T {\sin (\omega t)\langle \tilde \psi_+ |{\sigma _y}|\tilde \psi_+ \rangle d\tau }  + \frac{A}{2}\int_0^T {\cos (\omega t)\langle \tilde \psi_+ |{\sigma _x}|\tilde \psi_+ \rangle d\tau }.
\end{eqnarray}
We calculate the three integrals
\begin{subequations}
\begin{equation}
    \int_0^T {\langle \tilde \psi_+ |{\sigma _z}|\tilde \psi_+ \rangle d\tau }= \frac{1}{2}\frac{{2{{\tilde \delta }^2}\left( {{c_1}^2 - {c_2}^2} \right) + 2\tilde A\tilde \delta {c_1}{c_2}}}{{{{\tilde \Omega }^2}}}T+ \frac{1}{2}\frac{{{{\tilde A}^2}\left( {{c_1}^2 - {c_2}^2} \right) - 4\tilde A\tilde \delta {c_1}{c_2}}}{{2{{\tilde \Omega }^2}}}\frac{{\sin (\tilde \Omega T)}}{{\tilde \Omega }},
\end{equation}
\begin{eqnarray}
    \fl\int_0^T {2\sin (\omega \tau){\langle \tilde \psi_+ |{\sigma _z}|\tilde \psi_+ \rangle d\tau } }  = \frac{{{{\tilde A}^2}{c_1}{c_2} + \tilde A\tilde \delta \left( {{c_1}^2 - {c_2}^2} \right)}}{{2{{\tilde \Omega }^2}}}T\nonumber\\\fl
    + \frac{{4{{\tilde \delta }^2}{c_1}{c_2} - \tilde A\tilde \delta \left( {{c_1}^2 - {c_2}^2} \right)}}{{2{{\tilde \Omega }^2}}}\left( {\frac{{\sin (\tilde \Omega T)}}{{\tilde \Omega }} - \frac{{\tilde \Omega \sin (\tilde \Omega T)}}{{{{\tilde \Omega }^2} - 4{\omega ^2}}}} \right)\nonumber\\\fl
    - \frac{{-4\tilde \delta {c_1}{c_2} + \tilde A\left( {{c_1}^2 - {c_2}^2} \right)}}{{2\tilde \Omega }}\frac{{2\omega \sin (\tilde \Omega T)}}{{{{\tilde \Omega }^2} - 4{\omega ^2}}},
\end{eqnarray}
\begin{eqnarray}\fl
    \int_0^T {2\cos (\omega \tau){\langle \tilde \psi_+ |{\sigma _z}|\tilde \psi_+ \rangle d\tau } }  = \frac{{{{\tilde A}^2}{c_1}{c_2} + \tilde A\tilde \delta \left( {{c_1}^2 - {c_2}^2} \right)}}{{2{{\tilde \Omega }^2}}}T \nonumber\\\fl
    + \frac{{4{{\tilde \delta }^2}{c_1}{c_2} - \tilde A\tilde \delta \left( {{c_1}^2 - {c_2}^2} \right)}}{{2{{\tilde \Omega }^2}}}\left( {\frac{{\sin (\tilde \Omega T)}}{{\tilde \Omega }} + \frac{{\tilde \Omega \sin (\tilde \Omega T)}}{{{{\tilde \Omega }^2} - 4{\omega ^2}}}} \right) \nonumber\\\fl
    + \frac{{-4\tilde \delta {c_1}{c_2} + \tilde A\left( {{c_1}^2 - {c_2}^2} \right)}}{{2\tilde \Omega }}\frac{{2\omega \sin (\tilde \Omega T)}}{{{{\tilde \Omega }^2} - 4{\omega ^2}}},
\end{eqnarray}
\end{subequations}
separately. Thus, $\int_0^T {\langle \tilde \psi_+ |{e^S}H{e^{ - S}}|\tilde \psi_+ \rangle d\tau }$ reads as
\begin{eqnarray}\fl
    \int_0^T {\langle {{\tilde \psi }_ + }|{e^S}H{e^{ - S}}|{{\tilde \psi }_ + }\rangle d\tau } \nonumber\\\fl
    = \frac{{\tilde \Delta }}{2}\int_0^T {\langle {{\tilde \psi }_ + }|{\sigma _z}|{{\tilde \psi }_ + }\rangle d\tau } + \frac{{\tilde A}}{4}\int_0^T {\sin (\omega t)\langle \tilde \psi_+ |{\sigma _y}|\tilde \psi_+ \rangle d\tau } + \frac{A}{2}\int_0^T {\cos (\omega t)\langle \tilde \psi_+ |{\sigma _x}|\tilde \psi_+ \rangle d\tau } \nonumber\\\fl
    = \frac{{\tilde AT}}{{2{{\tilde \Omega }^2}}}\left( {  \frac{{\tilde \Delta \tilde \delta }}{{\tilde A}} + \frac{{\tilde A}}{8} + \frac{A}{4}} \right)\left[ {  \tilde \delta \left( {{c_1}^2 - {c_2}^2} \right) + \tilde A{c_1}{c_2}} \right]\nonumber\\\fl
    + \left( {\frac{{\tilde \Delta \tilde A}}{4} - \frac{{\tilde A\tilde \delta }}{8} - \frac{{A\tilde \delta }}{4}} \right)\frac{{\sin (\tilde \Omega T)}}{{2{{\tilde \Omega }^3}}}\left[ {\tilde A\left( {{c_1}^2 - {c_2}^2} \right) - 4\tilde \delta {c_1}{c_2}} \right]\nonumber\\\fl
    + \left( {\frac{A}{4} - \frac{{\tilde A}}{8}} \right)\frac{{\left( {-\tilde \delta  + 2\omega } \right)\sin (\tilde \Omega T)}}{{2\tilde \Omega \left( {{{\tilde \Omega }^2} - 4{\omega ^2}} \right)}}\left[{\tilde A\left( {{c_1}^2 - {c_2}^2} \right) - 4\tilde \delta {c_1}{c_2}} \right].
\label{A10}
\end{eqnarray}
In the 3rd harmonic resonance regime $\frac{\sin(\tilde\Omega T)}{2\tilde\Omega^3} \ll 1$ so we omit the second term. Substituting Eq. (\ref{xinchutai}) into (\ref{A10}), we get
\begin{eqnarray}\fl
   \int_0^T {\langle \tilde \psi_+ |{e^S}H{e^{ - S}}|\tilde \psi_+ \rangle d\tau }
   = &-& \frac{1}{{\sqrt {1 + {k^2}} }}\left( {  \frac{{\tilde \Delta \tilde \delta }}{{\tilde A}} + \frac{{\tilde A}}{8} + \frac{A}{4}} \right)\frac{{\tilde AT}}{{2\tilde \Omega }} \nonumber\\\fl
   &-& \frac{k}{{\sqrt {1 + {k^2}} }}\left( {\frac{A}{4} - \frac{{\tilde A}}{8}} \right)\frac{{-\tilde \delta  + 2\omega }}{{{{\tilde \Omega }^2} - 4{\omega ^2}}}\sin (\tilde \Omega T).
\end{eqnarray}
We finally obtain the geometric phases of Eq. (\ref{3HR}).

\section{Calculations near the 3rd harmonic resonance point\label{appendix2}}
All calculations in this part are performed under the condition of $\Delta\approx\Delta _{\rm res}$ and $\tilde\Omega\approx2\omega$. From \cite{CHRW}, we get $\xi\approx\frac{\omega }{{\omega  + \Delta }}$ and $Z=\frac{A}{\omega}\xi\approx\frac{A}{{\omega  + \Delta }}$. Bessel function $J_{n}(Z) (n=0,1,2,3)$ can be expanded in $Z$
\begin{subequations}
\begin{equation}
    {J_0}(Z) \approx 1 - {\left( {\frac{Z}{2}} \right)^2} = 1 - \frac{{{A^2}}}{{4{{\left( {\omega  + \Delta } \right)}^2}}},
    \label{J0}
\end{equation}
\begin{equation}
    {J_1}(Z) \approx \frac{Z}{2} = \frac{A}{{2\left( {\omega  + \Delta } \right)}},
    \label{J1}
\end{equation}
\begin{equation}
    {J_2}(Z) \approx \frac{1}{2}{\left( {\frac{Z}{2}} \right)^2} = \frac{1}{8}\frac{{{A^2}}}{{{{\left( {\omega  + \Delta } \right)}^2}}},
    \label{J2}
\end{equation}
\begin{equation}
    {J_3}(Z) \approx \frac{1}{6}{\left( {\frac{Z}{2}} \right)^3} = \frac{1}{{48}}\frac{{{A^3}}}{{{{\left( {\omega  + \Delta } \right)}^3}}}.
    \label{J3}
\end{equation}
\end{subequations}
Thus $\tilde{\Delta}$, $\tilde{\delta}$ and $\tilde{A}$ are obtained
\begin{subequations}
\begin{equation}
    \tilde \Delta  = \Delta {J_0}(Z) \approx \Delta  - \frac{{\Delta {A^2}}}{{4{{\left( {\omega  + \Delta } \right)}^2}}},
    \label{D_N}
\end{equation}
\begin{equation}
    \tilde \delta  =\tilde \Delta- \omega  \approx \Delta-\omega  - \frac{{\Delta {A^2}}}{{4{{\left( {\omega  + \Delta } \right)}^2}}},
    \label{d}
\end{equation}
\begin{equation}
    \tilde A = 4\Delta{J_1}(Z) \approx \frac{{2\Delta A}}{{\omega  + \Delta }},
    \label{A_N}
\end{equation}
\end{subequations}
respectively. From Eq. (\ref{point}), we define $\Delta _{\rm res}=(3-x)\omega$ where $x$ is a small quantity,
\begin{equation}
    x=\frac{3A^2}{32\omega^2}\label{x}.
\end{equation}


\subsection{Change rate of geometric phase with $\Delta/\omega$\label{appendix2.1}}
From Eq. (\ref{k}), $k$ can be approximately written as $k=\frac{k_0}{\tilde\Omega-2\omega}$ where $k_0$ is a constant
\begin{equation}
    k_0 = \frac{{\Delta \left[ {{J_2}\left( Z \right)\tilde A - {J_3}\left( Z \right)\left( {\tilde \Omega  + \tilde \delta } \right)} \right]}}{\tilde\Omega+2\omega}\ll1.
\end{equation}
By Eq. (\ref{J2}), (\ref{J3}), (\ref{d}) and (\ref{A_N}), $k_0$ can be calculated approximately
\begin{eqnarray}
    {k_0} &\approx& \frac{{{\Delta _{\rm res}}{A^3}\left[ {11{\Delta _{\rm res}} - \omega  + \frac{{{\Delta _{\rm res}}{A^2}}}{{4{{\left( {\omega  + {\Delta _{\rm res}}} \right)}^2}}}} \right]}}{{192\omega {{\left( {\omega  + {\Delta _{\rm res}}} \right)}^3}}} \nonumber\\
    &\approx& \frac{{{\Delta _{\rm res}}{A^3}\left( {11{\Delta _{\rm res}} - \omega } \right)}}{{192\omega {{\left( {\omega  + {\Delta _{\rm res}}} \right)}^3}}}.
    \label{k0}
\end{eqnarray}
The change of geometric phase mainly comes from second term in Eq. (\ref{3HR}) near the 3rd harmonic resonance point. From Eqs. (\ref{point}), (\ref{d}) and (\ref{A_N}), the second term in Eq. (\ref{3HR}) can be written as
\begin{eqnarray}
    \fl - \frac{1}{{\sqrt {1 + {k^2}} }}\left( {  \frac{{\tilde \Delta \tilde \delta }}{{\tilde A}} + \frac{{\tilde A}}{8} + \frac{A}{4}} \right)\frac{{\tilde A}}{{2\tilde\Omega }}T \nonumber\\
    \fl\approx  - \left( {\tilde \Omega  - 2\omega } \right)\frac{{48{{\left( {\omega  + \Delta _{\rm res} } \right)}^3}}}{{{A^3}\left( {11\Delta _{\rm res}  - \omega } \right)}}\frac{{4\left( {\Delta _{\rm res}  + \omega } \right)\left( {{\Delta _{\rm res} ^2} - {\omega ^2}} \right) + {A^2}\left( {2\Delta _{\rm res}  + 3\omega } \right)}}{{4{{\left( {\Delta _{\rm res}  + \omega } \right)}^2}}}T.
\end{eqnarray}

The change rate of $\tilde\Omega$ with $\Delta$ near the 3rd harmonic resonance point can be obtained
\begin{eqnarray}
    \frac{{d\tilde \Omega }}{{d\Delta }} &\approx& \frac{{2\left( {{\Delta _{\rm res}} - \omega } \right) + \frac{{{A^2}\omega }}{{2{{\left( {\omega  + {\Delta _{\rm res}}} \right)}^2}}}}}{{2\sqrt {{{\left( {\omega  - {\Delta _{\rm res}}} \right)}^2} + \frac{{{\Delta _{\rm res}}{A^2}}}{{2\left( {\omega  + {\Delta _{\rm res}}} \right)}}} }} \approx \frac{{{\Delta _{\rm res}} - \omega }}{{2\omega }}=1-\frac{x}{2}.
    \label{O/D}
\end{eqnarray}
The dimensionless change rate $|\frac{d\gamma}{d(\Delta/\omega)}|$ can be expanded to the leading order in $x$

\begin{eqnarray}\fl
    \left| {\frac{{d\gamma }}{{d(\Delta/\omega) }}} \right| \approx \left|\frac{d}{d(\Delta/\omega)}\left[- \frac{1}{{\sqrt {1 + {k^2}} }}\left( {  \frac{{\tilde \Delta \tilde \delta }}{{\tilde A}} + \frac{{\tilde A}}{8} + \frac{A}{4}} \right)\frac{{\tilde A}}{{2\tilde\Omega }}T\right] \right|\nonumber\\\fl
    \approx \left| \omega{\frac{{d\tilde \Omega }}{{d\Delta }}} \right|\left| {\frac{{48{{\left( {\omega  + {\Delta _{\rm res}}} \right)}^3}}}{{{A^3}\left( {11{\Delta _{\rm res}} - \omega } \right)}}\frac{{4\left( {{\Delta _{\rm res}} + \omega } \right)\left( {\Delta _{_{res}}^2 - {\omega ^2}} \right) + {A^2}\left( {2{\Delta _{\rm res}} + 3\omega } \right)}}{{4{{\left( {{\Delta _{\rm res}} + \omega } \right)}^2}}}T} \right|\nonumber\\\fl
    \approx \frac{{3\omega^2 }}{{2{A^3}}}\left[ {128{\omega ^2}\left( {1 - \frac{{45}}{{32}}x} \right) + 9{A^2}\left( {1 - \frac{{181}}{{288}}x} \right)} \right]T
    = 3\pi \left[ {128{{\left( {\frac{\omega }{A}} \right)}^3} - \frac{{63}}{8} {\frac{\omega }{A}}} \right].
\end{eqnarray}

We obtain Eq. (\ref{xielv}).


\subsection{Quasienergies\label{appendix2.2}}
From Eqs. (\ref{quasienergy}) and (\ref{O/D}), near the 3rd harmonic resonance point quasienergies $q_\pm$ can be written
\begin{eqnarray}
    q_ \pm &=&  \mp \frac{{\sqrt {k_0^2 + {{(\tilde \Omega  - 2\omega )}^2}} }}{2} - \frac{{\left( {2n + 1} \right)\omega }}{2} \nonumber\\
    &\approx&  \mp \frac{{{k_0} + \frac{{{{\left( {1 - \frac{x}{2}} \right)}^2}}}{{2{k_0}}}{{(\Delta  - {\Delta _{\rm res}})}^2}}}{2} + \frac{{\left( {2n + 1} \right)\omega }}{2}.
\end{eqnarray}
At the 3rd harmonic resonance point, $q_\pm$ can be obtained
\begin{equation}
    {q_ \pm } =  \mp \frac{{{k_0}}}{2} + \frac{{\left( {2n + 1} \right)\omega }}{2},
\end{equation}
where $k_0$ is the gap $\Xi$. Finally, by Eqs. (\ref{point}) and (\ref{k0}) we  obtain Eqs. (\ref{gap}) and (\ref{q zhengbi}).


\providecommand{\newblock}{}

\noappendix
\newcommand\x{0.3}
\clearpage
\begin{figure*}
\centering
\subfigure{
    \label{fig_A1}
    \includegraphics[width=\x\linewidth]{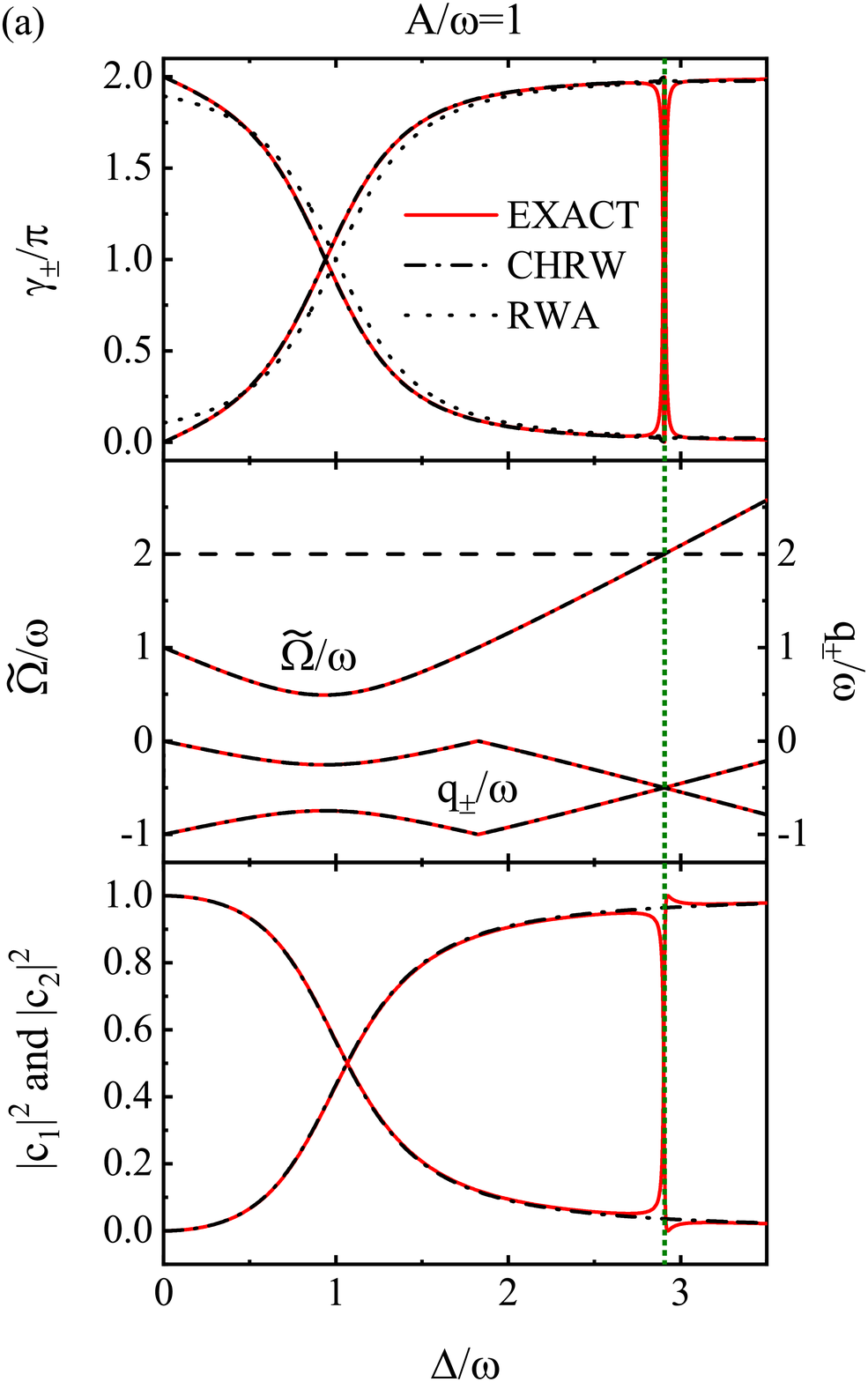}}
\subfigure{
    \label{fig_A2}
    \includegraphics[width=\x\linewidth]{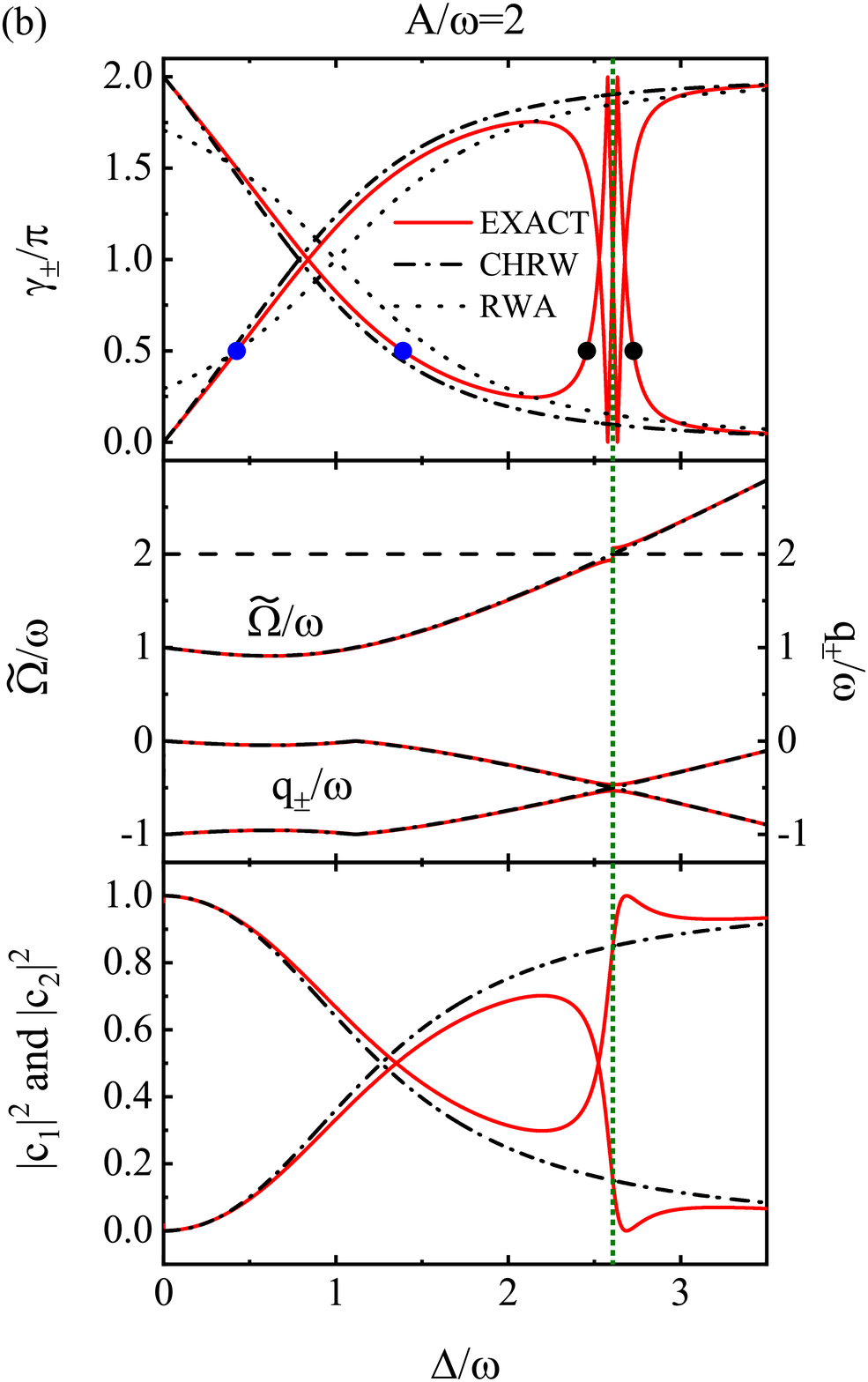}}
\caption{Geometric phases $\gamma_{\pm}$ (top pannel), Rabi frequency $\tilde{\Omega}/\omega$ and quasienergies $q_{\pm}$ (middel pannel), $|c_1|^2$ and $|c_2|^2$ of the cyclic initial state (bottom pannel) as a function of $\Delta/\omega$ for $A/\omega=1$ and $2$, which are shown in (a) and (b), respectively. The numerically exact results are plotted by the red line, the CHRW results by the dash-dotted line, the RWA result by the dotted line. In each figure, $\tilde\Omega/\omega=2$ is plotted by the black dashed line and $\Delta=\Delta _{\rm res}$ by the olive short dotted line. In the middle pannel, the line above shows the $\tilde{\Omega}/\omega$ and the two lines below show the quasienergies $q_{\pm}$. In (b), the interval of $\Delta/\omega$ between blue-point indicators is the main harmonic resonance regime while the interval of $\Delta/\omega$ between black-point indicators is the third harmonic resonance regime.}
\label{fig}
\end{figure*}


\clearpage
\begin{figure*}
\centering
\subfigure{
    \label{fig_A1PT}
    \includegraphics[width=\x\linewidth]{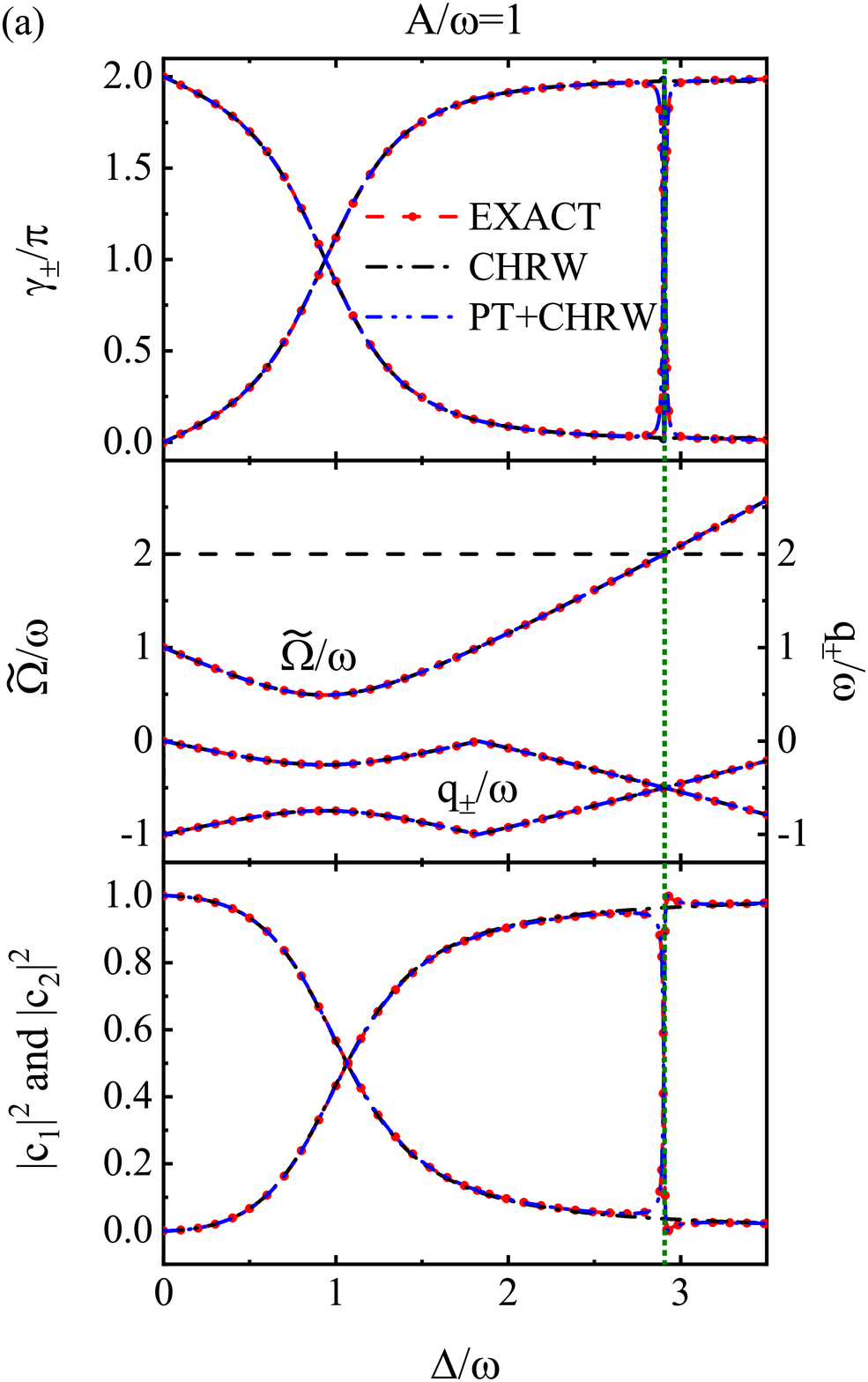}}
\subfigure{
    \label{fig_A2PT}
    \includegraphics[width=\x\linewidth]{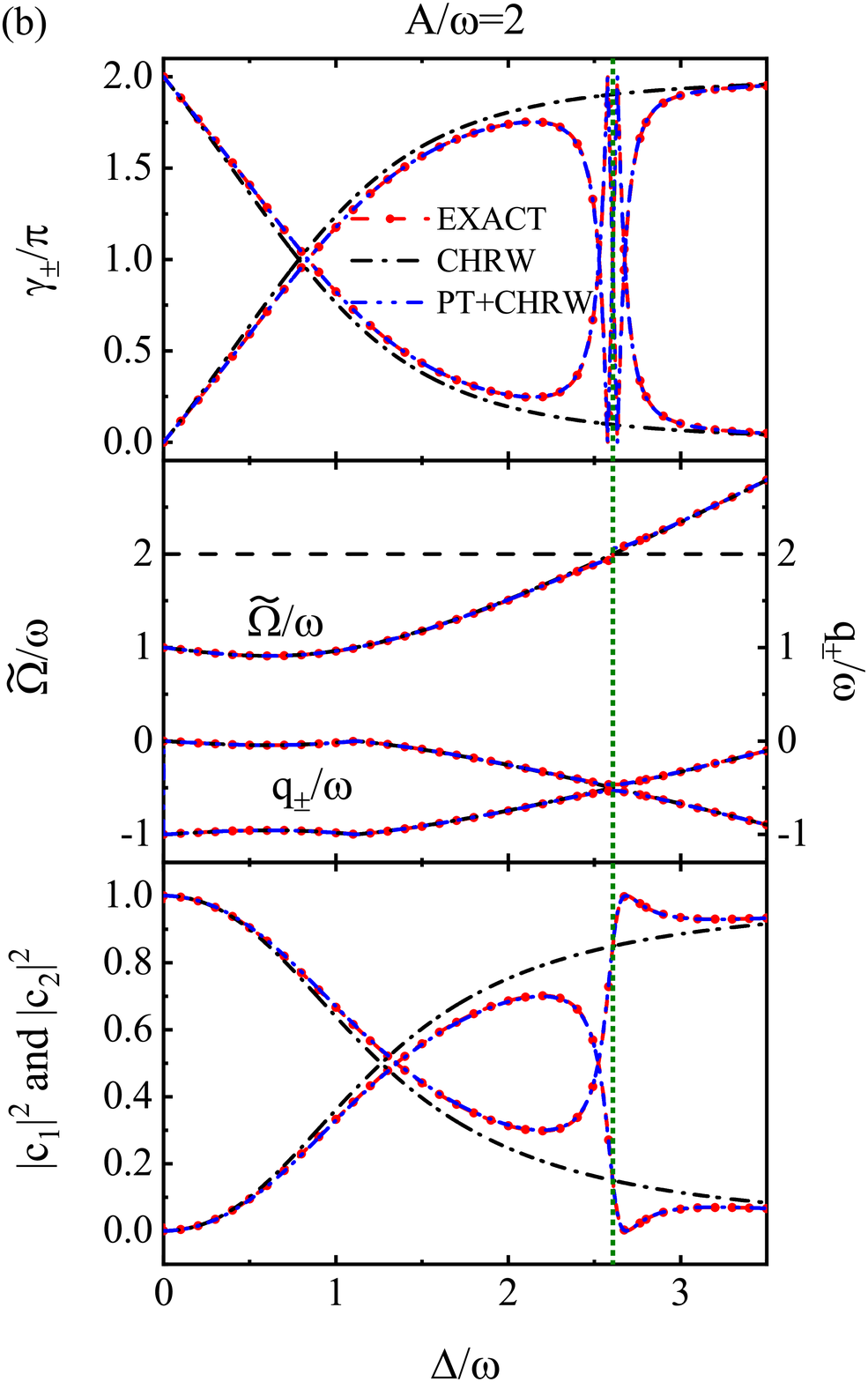}}

\subfigure{
    \label{fig_D3A1}
    \includegraphics[width=\x\linewidth]{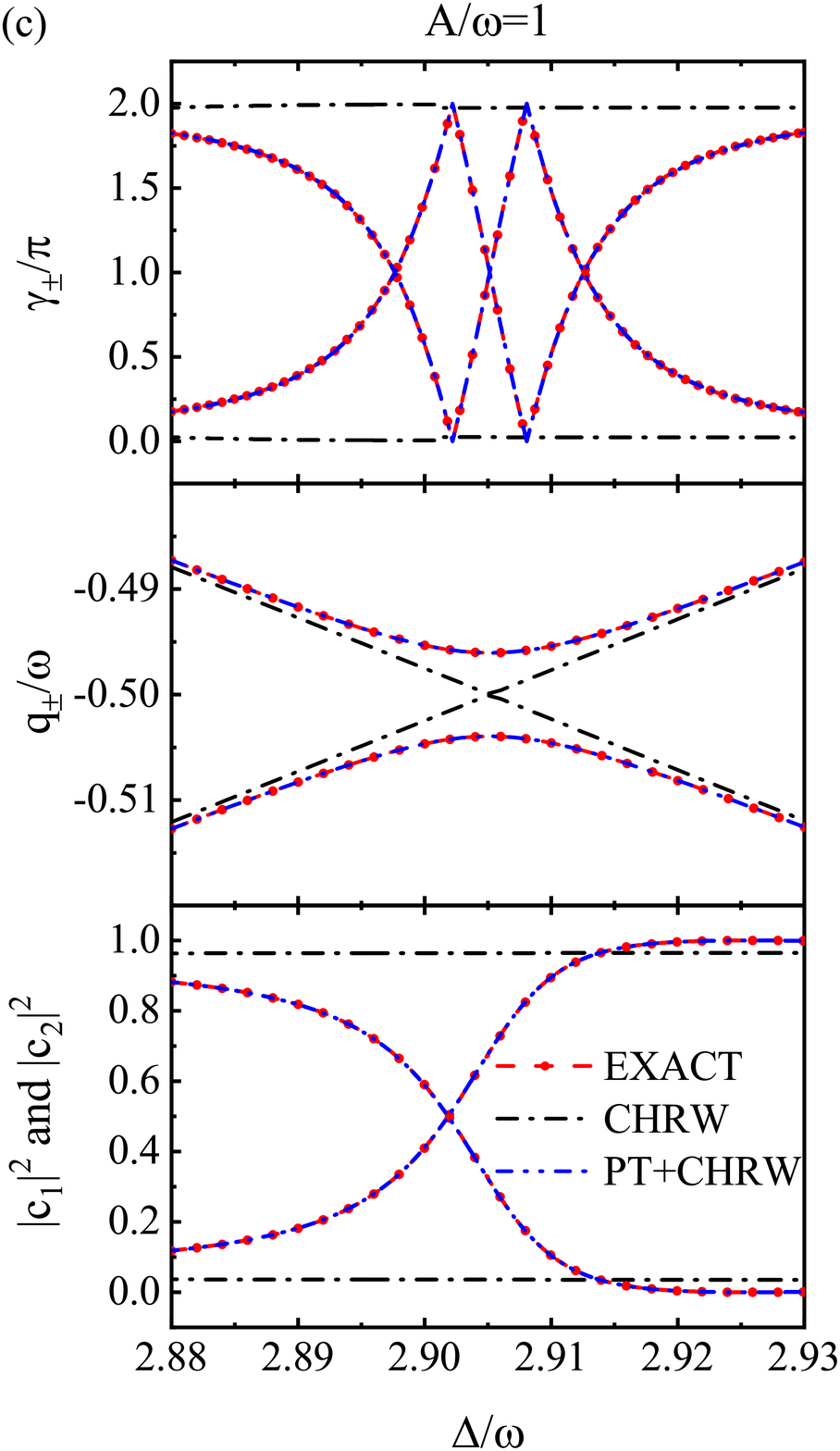}}
\subfigure{
    \label{fig_D3A2}
    \includegraphics[width=\x\linewidth]{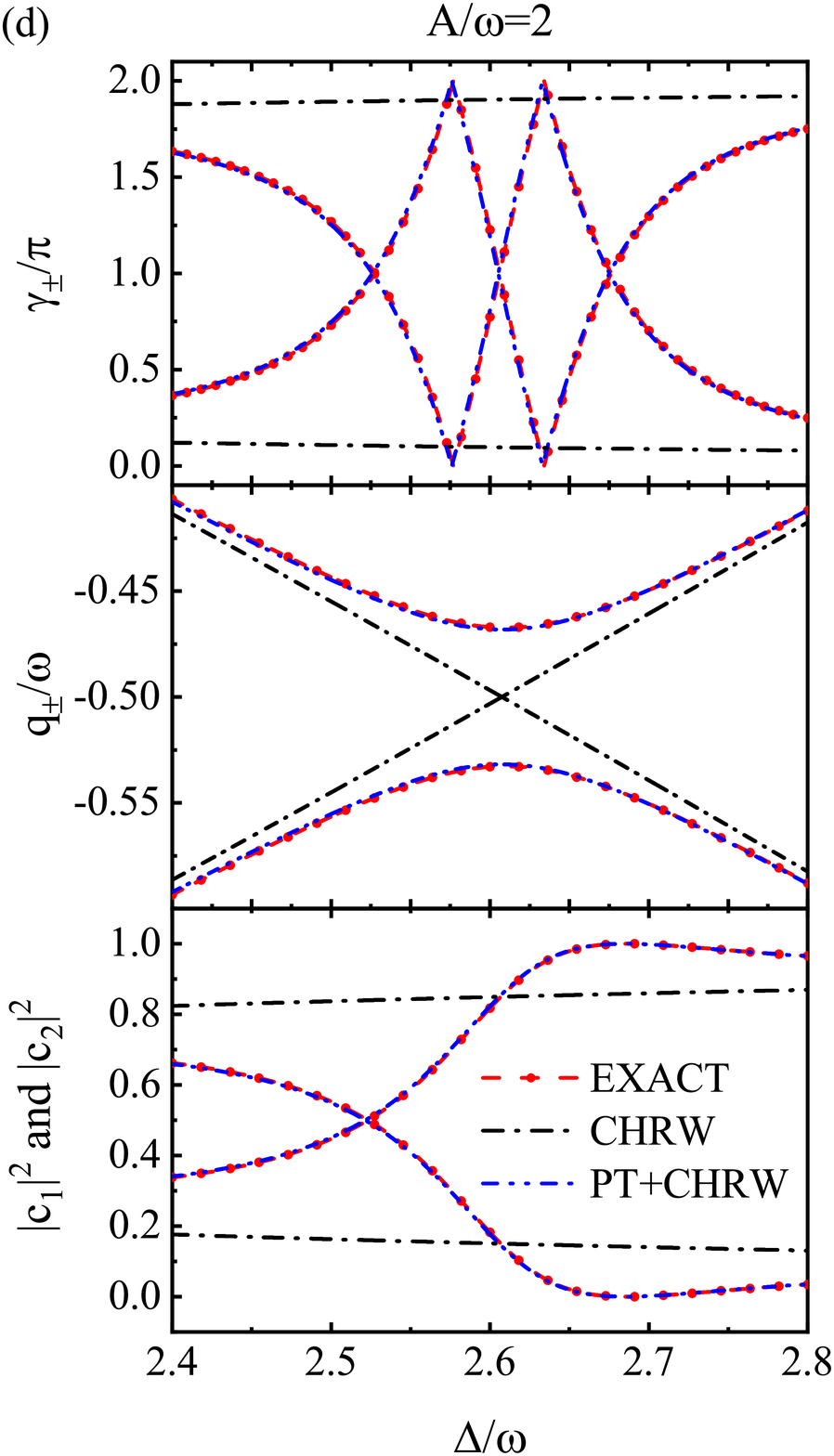}}
\caption{Geometric phases $\gamma_{\pm}$ (top pannel), Rabi frequency $\tilde{\Omega}/\omega$ and quasienergies $q_{\pm}$ (middel pannel), $|c_1|^2$ and $|c_2|^2$ of the cyclic initial state (bottom pannel) as a function of $\Delta/\omega$ for $A/\omega=1$ ((a) and (c)) and $A/\omega=2$ ((b) and (d)), respectively.  The numerically exact results are plotted by the red dashed line with filled cycles. PT+CHRW (shown with the blue dash-dot-dotted line) denotes the results of perturbation theory based on the CHRW method. In each figure, $\tilde\Omega/\omega=2$ is plotted by the black dashed line and $\Delta=\Delta _{\rm res}$ by the olive short dotted line. In the middle pannel of (a) and (b), the line above shows the $\tilde{\Omega}/\omega$ and the two lines below show the quasienergies $q_{\pm}$.}
\label{fig_PT}
\end{figure*}


\begin{figure*}
\centering
\subfigure{
    \label{fig_xielv}
    \includegraphics[height=5cm]{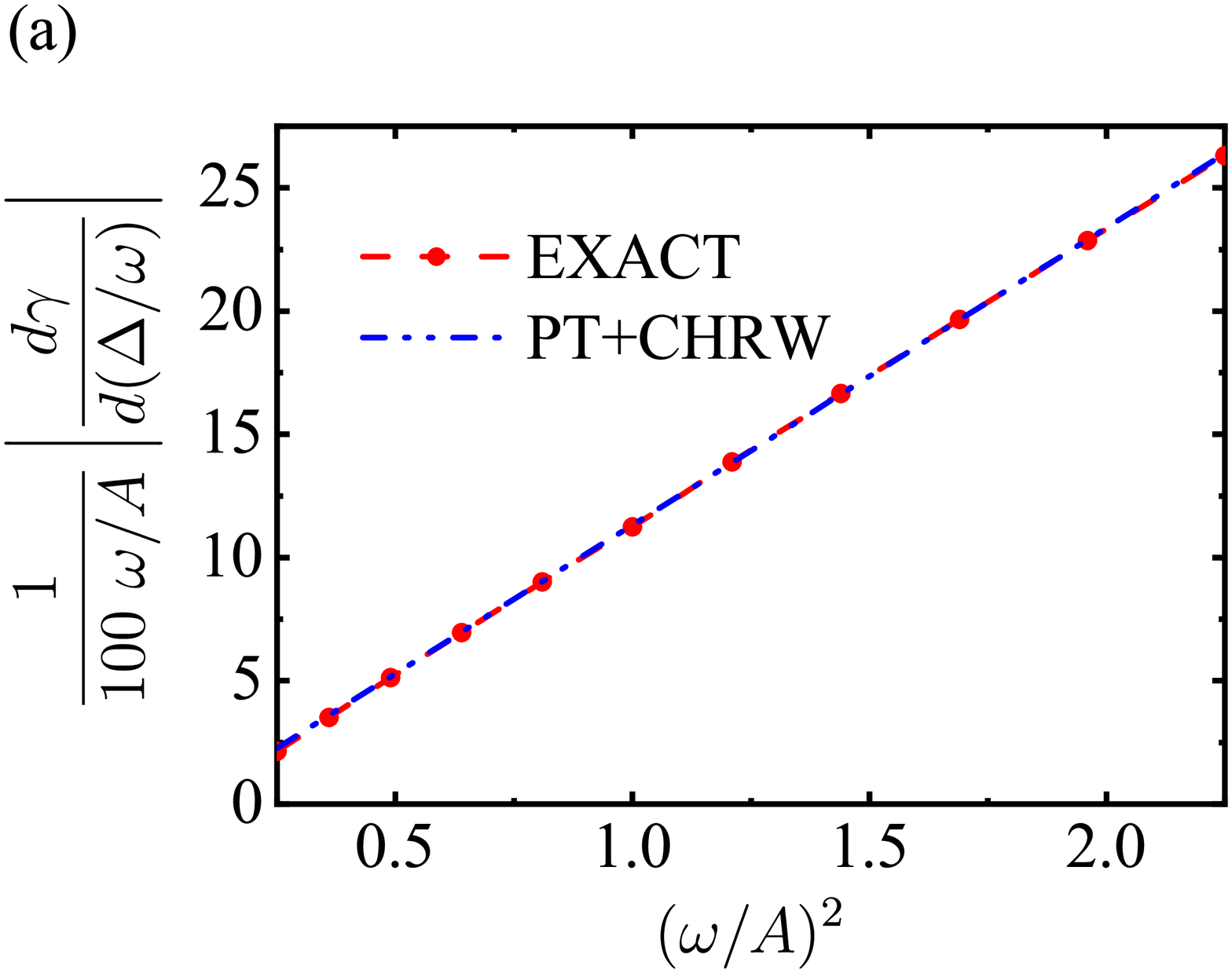}}
\subfigure{
    \label{fig_gap}
    \includegraphics[height=5cm]{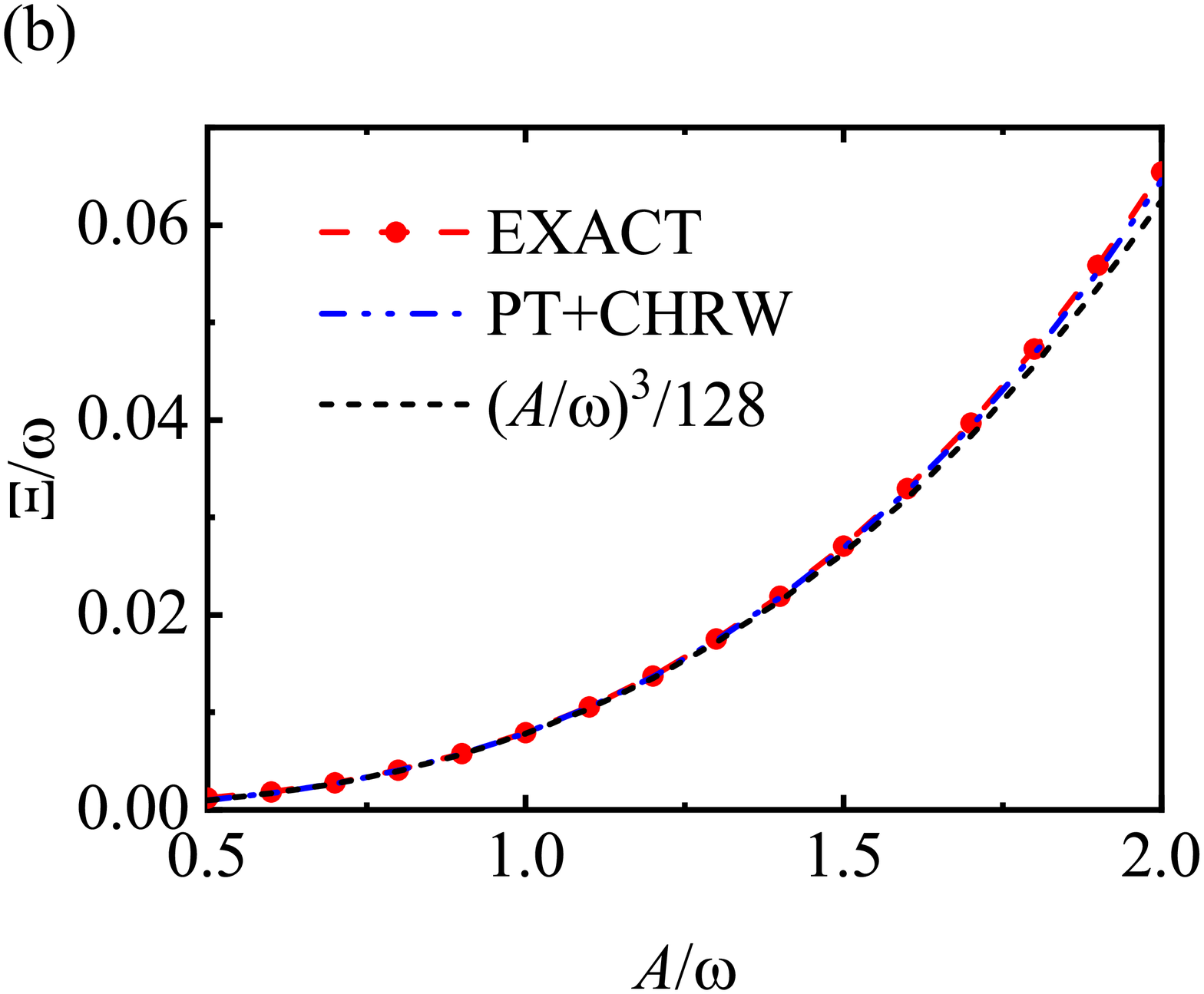}}
\caption{The properties of dramatic change of geometric phase in 3rd harmonic resonance regime. (a)$\frac{1}{100~\omega/A}\left|{\frac{d\gamma}{d(\Delta/\omega)}}\right|$ as a function of $(\omega/A)^2$.
(b) Dimensionless gap $\Xi/\omega$ between quasienergies at the 3rd harmonic resonance point as a function of $A/\omega$. The numerically exact results are plotted by the red dashed line with filled cycles. PT+CHRW (shown with the blue dash-dot-dotted line) denotes the results of perturbation theory based on the CHRW method.}
\end{figure*}

\newcommand\y{0.4}
\begin{figure*}
\centering
\subfigure{
    \label{5a}
    \includegraphics[width=\y\linewidth,trim={130 100 130 60},clip]{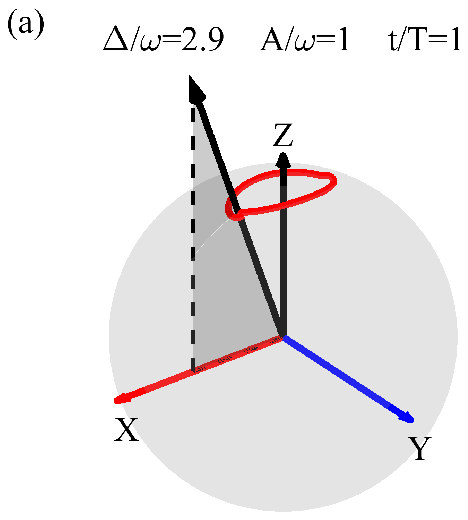}}
\subfigure{
    \label{5b}
    \includegraphics[width=\y\linewidth,trim={130 100 130 60},clip]{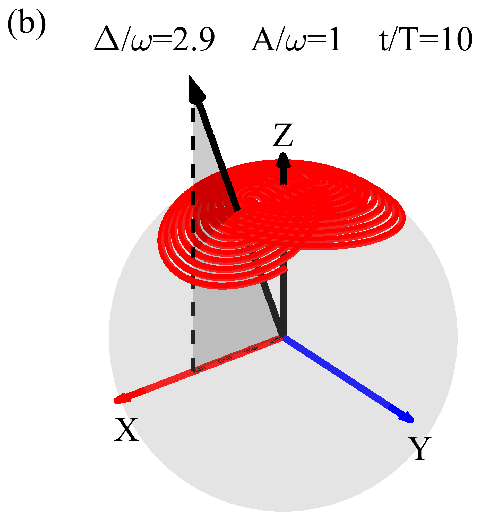}}

\subfigure{
    \label{5c}
    \includegraphics[width=\y\linewidth,trim={130 100 130 60},clip]{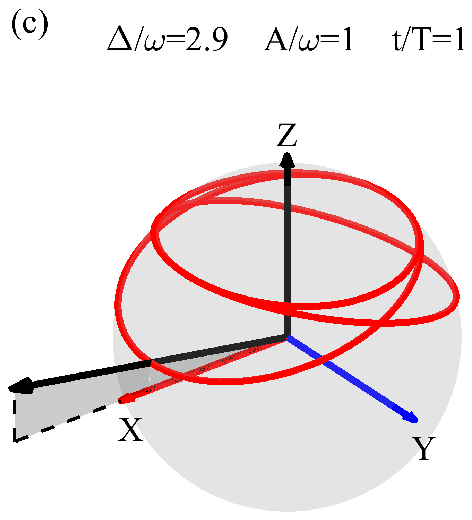}}
\subfigure{
    \label{5d}
    \includegraphics[width=\y\linewidth,trim={130 100 130 60},clip]{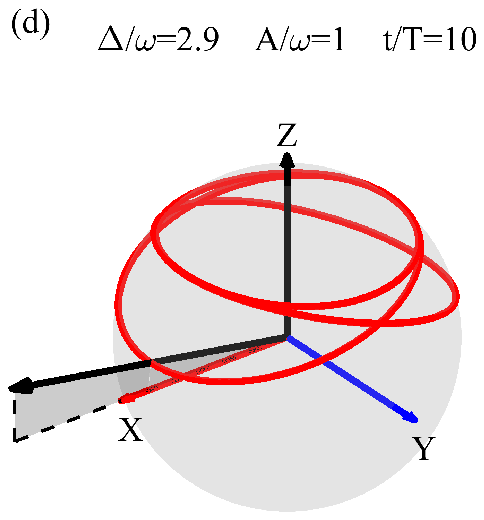}}
\caption{Time evolution of a qubit initially prepared as instantaneous eigenstate ((a) for $t/T=1$ and (b) for $t/T=10$) and cyclic initial state ((c) for $t/T=1$ and (d) for $t/T=10$) is shown on the Bloch sphere with $\Delta/\omega=2.9$ and $A/\omega=1$. The long solid line with an arrow denotes an initial state.}
\label{fig4}
\end{figure*}

\begin{figure}
\centering
\includegraphics[width=\x\linewidth]{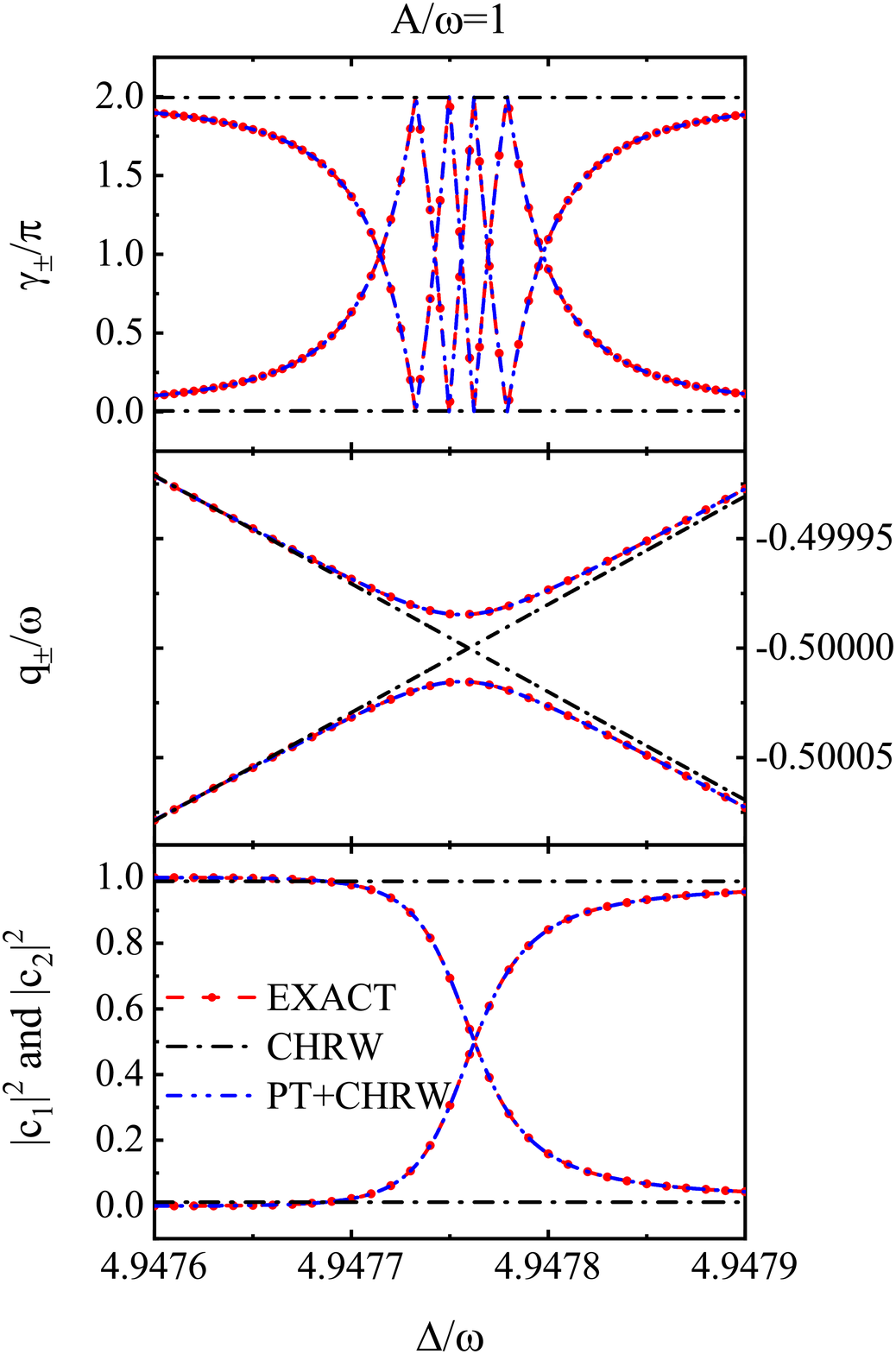}
\caption{Geometric phases $\gamma_{\pm}$ (top pannel), quasienergies (middel pannel), $|c_1|^2$ and $|c_2|^2$ of the cyclic initial state (bottom pannel) as a function of $\Delta/\omega$ for $A/\omega=1$ in the 5th harmonic resonance regime. The numerically exact results are plotted by the red dashed line with filled cycles. PT+CHRW (shown with the blue dash-dot-dotted line) denotes the results of perturbation theory based on the CHRW method.}
\label{fig_D5}
\end{figure}

\end{document}